\documentclass[preprint]{elsarticle}
\biboptions{numbers,sort&compress}
\usepackage{graphicx}
\usepackage[utf8]{inputenc}
\usepackage{blindtext}
\usepackage[scr=rsfs]{mathalpha}
\usepackage{pgfplots}
\usepackage[T1]{fontenc}
\usepackage[margin=1.0in]{geometry}
\usepackage{float}
\usepackage{fullpage}
\usepackage{amsmath}
\usepackage{array}
\usepackage{hyperref}
\usepackage{physics}
\usepackage{amsfonts}
\usepackage{lineno}
\usepackage{multirow}
\modulolinenumbers[10]
\numberwithin{equation}{section}
\usepackage{MnSymbol}
\usepackage{lipsum}
\usepackage{multicol}
\usepackage{subfigure}
\usepackage{multirow}
\usepackage{xcolor}

\pgfplotsset{compat=1.15}

\usepackage{subfigure} 
\makeatletter
\def\ps@pprintTitle{%
  \let\@oddhead\@empty
  \let\@evenhead\@empty
  \let\@oddfoot\@empty
  \let\@evenfoot\@oddfoot
}

\newcommand{\bm}[1]{\ensuremath{\mathbf{#1}}}
\newcommand{\bs}[1]{\ensuremath{\boldsymbol{#1}}}

\makeatother
\begin{document}


\title{Reconstructing Blood Flow in Data-Poor Regimes: A Vasculature Network Kernel for Gaussian Process Regression}

\author[1]{Shaghayegh Z. Ashtiani}
\author[2]{Mohammad Sarabian}
\author[3]{Kaveh Laksari}
\author[1]{Hessam Babaee\corref{cor1}}

\address[1]{Department of Mechanical Engineering and Materials Science, University of Pittsburgh, Pittsburgh, PA.}
\address[2]{W. L. Gore \& Associates Inc., Phoenix, AZ.}
\address[3]{Department of Mechanical Engineering, University of California Riverside, Riverside, CA.}
\ead{h.babaee@pitt.edu}

\begin{abstract}
 Blood flow reconstruction in the vasculature is important for many clinical applications. However, in clinical settings, the available data are often quite limited. For instance, Transcranial Doppler ultrasound (TCD) is a noninvasive clinical tool that is commonly used in the clinical settings to measure blood velocity waveform at several locations on brain's vasculature. This amount of data is grossly insufficient for training machine learning surrogate models, such as deep neural networks or Gaussian process regression. In this work, we propose a Gaussian process regression approach based on physics-informed kernels, enabling near-real-time reconstruction of blood flow in data-poor regimes. We introduce a novel methodology to reconstruct the kernel within the vascular network, which is a non-Euclidean space. The proposed kernel encodes both spatiotemporal and vessel-to-vessel correlations, thus enabling blood flow reconstruction in vessels that lack direct measurements. We demonstrate that any prediction made with the proposed kernel satisfies the conservation of mass principle. The kernel is constructed by running stochastic one-dimensional blood flow simulations, where the stochasticity captures the epistemic uncertainties, such as lack of knowledge about boundary conditions and uncertainties in vasculature geometries. We demonstrate the performance of the model on three test cases, namely, a simple Y-shaped bifurcation, abdominal aorta, and the Circle of Willis in the brain.

\end{abstract}



\maketitle
\section{Introduction}\label{sec:Intro}
Accurate blood flow reconstruction in the vasculature is vital for many clinical applications. For instance, precise blood flow analysis in cerebral vessels helps in diagnosing and monitoring conditions such as cerebral vasospasm, Moyamoya disease, and brain tumors. Accurate flow reconstruction in the aorta aids in evaluating aortic aneurysms, dissections, and coarctations, enabling timely interventions. Additionally, detailed blood flow assessment is crucial for diagnosing and managing conditions like coronary artery disease, heart failure, and peripheral artery disease \cite{ruedinger20214d}. These applications underscore the importance of accurate blood flow reconstruction, highlighting its extensive impact across various domains of cardiovascular medicine.

Although advanced computational methods, such as image-based models, can provide detailed predictions of spatiotemporal events, integrating them into clinical practice is still challenging. This limitation primarily arises from two key factors: modeling uncertainties, including the lack of knowledge of the subject-specific boundary conditions, and the substantial computational time required for their execution. In these models, computational fluid dynamics (CFD) techniques are implemented on the extracted geometry from imaging data such as computed tomography (CT) or magnetic resonance imaging (MRI) \cite{deshpande2021automatic,deshpande2022novel}. There exists a large body of literature on this type of blood flow modeling (for example see \cite{updegrove2017simvascular,milner1998hemodynamics,moore1999accuracy,macraild2024accelerated}) as well as one-dimensional (1D) CFD \cite{stergiopulos1992computer,westerhof2009arterial,lin2016characteristic,pahlevan2011physiologically}. Among image-based models Ref. \cite{taylor2013computational} is the first FDA-approved CFD simulation for coronary stenosis assessment.

Building regression models for blood flow properties in the vasculature is challenging due to insufficient clinical data. This is exemplified by techniques such as Doppler ultrasonography, which, although widely used, often suffer from poor spatial resolution, and generally does not provide measurements for every vessel.  To better illustrate the challenges of constructing regression models for blood flow, consider a naive linear regression model in the form of 
$u(x,t) = \sum_{i=1}^{n_x}\sum_{j=1}^{n_t} w_{i,j} \psi_i(x) \chi_j(t)$,
where \(\psi_i(x)\) and \(\chi_j(t)\) represent 1D spatial and temporal basis functions, respectively. With \(n_x, n_t \sim \mathcal{O}(50-100)\) basis functions required in each vessel, training such models necessitates over \(\mathcal{O}(10^3-10^4)\) measurements in each vessel to compute the weight coefficients (\(w_{i,j}\)), far exceeding what is typically available in clinical settings.

The underlying issue with linear regression models is that the basis functions are fixed and chosen before analyzing the data. Deep neural networks (DNNs) can be interpreted as \emph{adaptive} basis function approximations, where the basis functions are learned from the data. For example, a feed-forward neural network regression model with \(x\) and \(t\) as input units can be written as 
$u(x,t) = \sum_{i=1}^{r} w_i \phi_i(x,t;\theta)$,
where \(w_i\) are the weights of the last layer, \(\theta\) represents the vector of neural network parameters, and \(\phi_i(x,t;\theta)\) can be viewed as adaptive basis functions that are learned from the data, with \(r \ll n_x n_t\). Consequently, the total number of parameters in a DNN that need to be inferred is significantly smaller than in a naive linear regression model, which, to some extent, relaxes the amount of data required to train such models. However, the available data is still grossly insufficient for training such models. For example, the DNN cannot be trained for vessels for which no measurements are available, and for vessels with few measurements, this can result in overfit models.

To address the issue of training DNN in data-poor regimes, physics-informed neural networks (PINNs) have been utilized \cite{raissi2019physics}, in which the fluid dynamics conservation laws alongside clinical measurements \cite{kissas2020machine,sarabian2022physics,arzani2021uncovering,raissi2020hidden} are used to train the DNN.  Training PINNs for 1D blood flow models in the vasculature requires enforcing the conservation of mass and momentum in each vessel, as well as enforcing these constraints at vessel junctions. Specifically, a DNN is considered for each vessel. As a result, PINNs can overcome the issue of data scarcity by enforcing physical laws as regularizers to avoid model overfitting and estimate blood properties in vessels with no measurements. 

The computational cost of training PINNs, especially for large vasculature networks, can be significant. For instance, training the PINNs to model brain hemodynamics requires approximately 40 hours using a single NVIDIA Tesla T4 GPU card, starting from a random network initialization. In clinical settings, particularly for stroke, there is an urgent need for near-real-time models capable of rapidly processing and analyzing complex medical data. This enables healthcare professionals to make swift, informed decisions about patient care, leading to improved outcomes and reduced mortality rates. Accurate and timely modeling can help identify stroke subtypes, predict treatment responses, and optimize personalized treatment strategies, ultimately enhancing patient recovery and quality of life \cite{garcia2021predictive}. Another challenge in training PINNs for blood flow is the lack of direct area measurements for vessel cross-sectional area or pressure, which, as shown in \cite{sarabian2022physics}, necessitates first training another DNN to relate velocity to pressure before training the PINN, which further adds to the training cost. 


Bayesian regression techniques offer some favorable features that can be very important when building regression models in data-poor regimes.  Bayesian regression models provide posterior probability density distributions for quantities of interest as opposed to deterministic models that yield point estimates. The probabilistic predictions encode the epistemic uncertainties due to insufficient data. These uncertainty estimates can also be used for targeted data acquisition.    Gaussian processes (GPs) are powerful Bayesian regression models that have an analytical workflow and as a result, they can be trained quickly, especially in data-poor regimes, where the size of training data is small \cite{rasmussen2006gaussian}. Multi-fidelity models based on GPs have also proven useful for building regression models in data-poor regimes \cite{perdikaris2016model,forrester2008engineering,meng2021multi,costabal2019machine,babaee2018multi,fleeter2020multilevel,babaee2016multi,raissi2019parametric}. The multi-fidelity models enable the fusion of a small number of high-fidelity data points, for example, clinical measurements, with a large number of low-fidelity measurements for example obtained from CFD simulations. Since the developments of this work are related to GPs, we highlight two major challenges below in applying standard GP for blood flow regression in data-poor regimes:
\begin{enumerate}
    \item \textbf{Choice of kernel:} The choice of the kernel can significantly impact the accuracy of GP predictions, especially in data-poor regimes. Typically square exponential kernel or Matern kernel are used for GP. However, training the GP using these kernels requires a significant amount of data.  Determining a suitable kernel for blood flow prediction is not clear. Also, finding kernel hyperparameters requires solving a non-convex optimization problem. However, when the training size is very low, solving the optimization problem leads to a poor choice of hyperparameters and it can result in overfitting.  

     \item \textbf{GP for a vasculature network:} Although GPs can be used to learn functions in a Euclidean input space ($x$), i.e., input spaces that are subsets of $\mathbb{R}^d$, there is no standard procedure for applying GPs to a vasculature network, which can be mathematically represented as a graph. The underlying difficulty arises because the notion of distance between two points in Euclidean space should not be directly applied to the network domain. For example, two vessels that are very close to each other in space can have completely different velocities. A brute-force application of GP to vasculature would require one GP per vessel, with the input space for each vessel being Euclidean. This approach would ignore the vessel-to-vessel correlations. In such cases, it is not possible to train a GP for vessels without data. Additionally, this approach can result in overfitting for vessels with very few measurements. Moreover, it is not clear how the conservation of mass and momentum can be enforced at the junctions. Another potential solution is to embed the 1D vasculature network in a three-dimensional (3D) space. However, this approach has its own challenges. Vessels that are physically close to each other may exhibit completely different flow properties. This makes it difficult to find a kernel capable of learning smooth flow behavior along a vessel, where $x$ varies, but also of learning nearby vessels with distinctly different flows.
\end{enumerate}

Solutions have been proposed to address the aforementioned challenges.   Related to the choice of kernels different techniques have been proposed in which the kernel is learned from data; see for example  \cite{pang2019neural,neal2012bayesian,williams1996computing,lee2017deep,yang2019physics,yang2021physics}, in which the kernel is parameterized as a neural network.  In \cite{owhadi2019kernel}, kernel flow was introduced in which kernels are learned from data.  Multi-fidelity GP is also employed \cite{perdikaris2017nonlinear,raissi2017inferring,raissi2016deep} for problems that do not have enough high-fidelity data. Multi-fidelity GP compensates for the lack of high-fidelity data by incorporating low-fidelity data.   There is a lack of studies utilizing a global GP for a vasculature network. However, GP has been applied to vasculature-related problems such as uncertainty quantification of 3D in-stent restenosis \cite{ye2022uncertainty}, uncertainty quantification in a 1D fluid-dynamics model of the pulmonary circulation \cite{paun2021markov}, and predicting abdominal aortic aneurysm growth \cite{do2018prediction}.

In this paper, we introduce a methodology that utilizes Gaussian Process regression to reconstruct blood flow within a vasculature network, relying on a relatively small number of measurements. Specifically, we develop a spatiotemporal kernel that encompasses the entire vasculature, encoding the correlations both within and between vessels. The offline computational cost of constructing the kernel involves performing $\mathcal{O}(100)$ 1D simulations, which can be run concurrently and take only minutes. The online cost of performing inference is negligible. We demonstrate that the presented methodology can reconstruct blood flow velocity in a vasculature network consisting of 
$\mathcal{O}(10)$ vessels with only time series measurements at one or two points.

We first introduce our methodology in detail in Sections \ref{sec:GP} and \ref{sec:Method}. We then apply the methodology to three demonstrative problems with different levels of complexity and noise: 1) a Y-shaped bifurcation to develop the fundamentals for blood flow simulation in a simple geometry, 2) simulated blood flow in abdominal aorta, with a more complex geometry, and 3) blood flow inside brain's vasculature (Circle of Willis), with a high level of geometric complexity and using real-world experimental measurements (Section \ref{sec:problem}). Finally, in Section \ref{sec:Disc} we summarize our work, and observations, and discuss potential future work.

\section{Gaussian Process Regression}\label{sec:GP}

Since our work is based on the GP, we briefly review GP and we also introduce the notation that will be used in our methodology. Let $\mathbf{y} := [y_1, y_2, \ldots, y_N]^T \in \mathbb{R}^N$ denote the observed data corresponding to inputs $\mathbf{x} := [x_1, x_2, \ldots, x_N]^T \in \mathbb{R}^N$, where $N$ is the number of training points. The objective is to train a surrogate function $f(x)$ using the available data to perform predictions at any $x$. Using the GP framework, function $f(x)$ is approximated as follows \cite{rasmussen2006gaussian}:
\begin{equation}
    y=f(x) + \varepsilon,  \quad f(\mathbf{x})\sim \mathcal{G P}\left(0, k\left(x, x^{\prime} ; \boldsymbol \theta\right)\right),
    \label{GP1ex}
   \end{equation}
where $\varepsilon$ represents the noise with a zero-mean normal distribution  $\varepsilon \sim \mathcal{N} (0, \sigma^2_n )$ where $\sigma_n$ is the standard deviation,  $k(x, x'; \boldsymbol \theta)$ is the kernel function, and $\boldsymbol{\theta}$ is vector of hyperparameters.

Applying the GP model given by Eq.~\ref{GP1ex} to a discrete data points given by $(\mathbf x, \mathbf y)$ results in the following  normal distribution:
    \begin{equation}
       \mathbf{y} \sim \mathcal{N} (0, \mathbf{K}+ \sigma^2_n \mathbf{I}), \quad \mathbf{K}= k(\mathbf{x}, \mathbf{x}^{\prime} ; \boldsymbol{\theta}).
    \label{GP3ex}
   \end{equation}
   where $\mathbf{I} \in \mathbb{R}^{N \times N}$ is the identity matrix.  The kernel hyperparameters and the noise variance \( \sigma^2_n \) can be learned from the data by minimizing the negative logarithm of the marginal likelihood (NLML) given by:
    \begin{equation}
NLML(\theta)=\frac{1}{2} \mathbf{y}^T (\mathbf{K}+ \sigma^2_n \mathbf{I})^{-1} \mathbf{y}+\frac{1}{2} \ln |\mathbf{K}+ \sigma^2_n \mathbf{I}|+\frac{N}{2} \ln (2 \pi).
    \label{GP4ex}
   \end{equation}
Therefore, using the computed hyperparameters and $\sigma_n$, allows prediction for a new input $f(x^*)$ using its conditional distribution as follows:
\begin{equation}
f\left(x^*\right) \mid \mathbf{y} \sim \mathcal{N}\left(k\left(x^*, \mathbf{x}\right) (\mathbf{K}+ \sigma^2_n \mathbf{I})^{-1} \mathbf{y}, k\left(x^*, x^*\right)-k\left(x^*, \mathbf{x}\right) (\mathbf{K}+ \sigma^2_n \mathbf{I})^{-1} k\left(\mathbf{x}, x^*\right)\right) .
    \label{GP5ex}
\end{equation}
The posterior mean and variance are therefore:
    \begin{subequations}
\begin{equation}
\label{mean}
    \mu(x) = k\left(x, \mathbf{x}\right) (\mathbf{K}+ \sigma^2_n \mathbf{I})^{-1} \mathbf{y},
\end{equation}
\begin{equation}
\label{covariance}
    \sigma^2(x) =  k\left(x, x\right)-k\left(x, \mathbf{x}\right) (\mathbf{K}+ \sigma^2_n \mathbf{I})^{-1} k\left(\mathbf{x}, x\right),
\end{equation}
    \end{subequations}
where  $\mu(x)$ represents the GP  prediction and  $\sigma(x)$  is the posterior standard deviation that quantifies uncertainty around predictions.

\section{Methodology}\label{sec:Method}

To address the limitations associated with the choice of kernel for a vasculature network, we propose an algorithm to construct physics-informed kernels using data generated from a 1D blood flow model. This kernel captures both spatiotemporal and vessel-to-vessel correlations.

\subsection{A Global Spatiotemporal Model}
 We aim to build a surrogate model for the blood flow velocity in a vasculature network using only a few measurements. In the following, we adopt a \emph{weight-space view} of linear Bayesian regression. We denote the quantity of interest at vessel $(k)$ with $f^{(k)}(x,t): \Omega^{(k)} \times [0,T] \rightarrow \mathbb{R}, k=1,2,\dots, K$ over the inputs $(x,t)$, where  $\Omega^{(k)}$ denotes the 1D spatial domain along the centerline of vessel $k$ and $T$ in the time duration and $K$ is the number of vessels.   The function $f$ is the target quantity, which is the area-averaged blood flow velocity.    Consider the following model for learning $f^{(k)}(x,t)$:
 \begin{equation}
    f^{(k)}(x,t)=\sum_{i=1}^r w^{(k)}_i(t) \phi^{(k)}_i (x), \quad x \in \Omega^{(k)}, \quad t \in[0,T], \quad k=1,2, \dots K,
    \label{model-1}
   \end{equation}
   where $ \phi^{(k)}_i(x): \Omega^{(k)}  \rightarrow \mathbb{R} $ are \emph{local} spatial basis functions that are chosen a priori and   $w^{(k)}_i(t)$ are the time-dependent weights that must be learned from measurements. In this context, 'local' refers to basis functions that are defined within each vessel. As $r \rightarrow \infty$, the above model approximates a GP with a spatial kernel. In this model, either the weights have to be parameterized separately or they have to be learned at each instant of time. In that case,  the model requires training $K$, GP models at any given time—one GP for each vessel. Consequently, for the vessels with insufficient data, the model's predictions are poor. Also, since the model must be trained at each instant of time, its predictions will be poor at time instants for which we have insufficient data. The latter issue can be circumvented by using a model with \emph{spatiotemporal} basis functions as shown below:
\begin{equation}
    f^{(k)}(x,t)=\sum_{i=1}^r w^{(k)}_i \phi^{(k)}_i (x,t), \quad x \in \Omega^{(k)}, \quad t \in[0,T], \quad k=1,2, \dots K,
    \label{model-2}
   \end{equation}
   where  $\phi^{(k)}_i(x,t): \Omega^{(k)} \times [0,T] \rightarrow \mathbb{R}$ are spatiotemporal local basis functions and $[w^{(k)}_1, w^{(k)}_2, \dots, w^{(k)}_r]$ are the corresponding time-invariant weights. In the model given by Eq.~\ref{model-2}, the spatiotemporal correlations are encoded in the basis function \(\phi^{(k)}_i(x,t)\), and the weights can be learned from disparate spatiotemporal measurements. As a result, training the model described by Eq.~\ref{model-2} requires a significantly smaller amount of data compared to the model outlined in Eq.~\ref{model-1}. However, this model still utilizes local basis functions, and one model must be trained for each vessel. Therefore, for vessels with insufficient measurements, the model described by Eq.~\ref{model-2} can result in poor predictions.

   In the following, we present a model based on \emph{global} spatiotemporal basis functions that exploits spatiotemporal as well as vessel-to-vessel correlations. The model is described by:
   \begin{equation}
    f(x,t)=\sum_{i=1}^r w_i \phi_i (x,t), \quad x \in \Omega= \bigcup_{k=1}^K \Omega^{(k)}, \quad t \in[0,T],
    \label{model-3}
   \end{equation}
   where $\Omega$ denotes the global vasculature network, mathematically represented as the union of all vessels.  In the above model, $\phi_i(x,t): \Omega \times [0,T] \rightarrow \mathbb{R}$ are global spatiotemporal basis functions.  Note that in the above model, the weights ${w_i}$ are global coefficients; in other words, they are not vessel-dependent.

\subsection{Uncertainty Modeling via Stochastic Simulations} 
The performance of the model given by Eq. \ref{model-3} critically depends on the choice of the basis functions $\phi_i(x,t)$. The choice of the basis function in the weight-space view is closely related to choosing a kernel in the function-space view. Our approach uses stochastic simulations to construct the kernel, which has been employed in building empirical kernels; see, for example, \cite{yang2019physics,yang2021physics}. Previous developments have concentrated on Euclidean input spaces, i.e., the input space is a subset of $\mathbb{R}^d$. Moreover, these techniques require the explicit storage of the kernel matrix. The present work extends these developments to vasculature networks and addresses the challenge of explicit kernel storage, which is cost-prohibitive in terms of memory requirements for the current application.

In the following, we present a data-driven methodology to build global spatiotemporal basis functions using a stochastic 1D blood flow model.  The steps are explained below. Consider the 1D stochastic blood flow model expressed as:
\begin{equation}
\frac{\partial \bm q}{\partial t} = N(\bm q,t; \boldsymbol \xi),
\end{equation}
augmented with appropriate boundary and initial conditions. In the equation above, $\bm q = \bm q(x,t; \boldsymbol \xi)$ represents the state variable vector $\bm q = [u, A]$, where $u=u(x,t; \boldsymbol \xi)$ denotes the area-averaged velocity in each vessel,  $A=A(x,t; \boldsymbol \xi)$ represents the cross-sectional area, and $\boldsymbol \xi =[\xi_1, \xi_2, \dots, \xi_d]$ represents the random parameters. These parameters are defined based on \emph{epistemic uncertainties} --- uncertainties stemming from incomplete knowledge. For instance, exact inlet boundary conditions might be unknown in clinical settings. Similarly, the measurement of vessel areas, often derived from CT imaging, is subject to variability based on the image segmentation techniques employed. Another source of uncertainty in modeling 1D blood flow arises from the assumptions made about outflow boundary conditions. All these epistemic uncertainties are encapsulated within $\boldsymbol \xi$, with $d$ representing the total number of these uncertain parameters. 

Given the solution of the 1D model, the spatiotemporal global correlation operator can be formed as shown below:
\begin{equation}
\mathcal{K}(x,t,x',t') = \mathbb{E}[u(x,t;\bs \xi)u(x',t';\bs \xi)], \quad x, x' \in  \Omega,  \quad \mbox{and} \quad t, t' \in [0,T].
\end{equation}
where $\mathbb{E}[\sim]$ is the expectation operator defined as:
\begin{equation}
    \mathbb{E}[u(x,t; \bs \xi)] = \int_{\mathbb{R}^d} u(x,t; \bs \xi) \rho_u(\bs \xi) d\bs \xi, 
\end{equation}
where $\rho_u(\bs \xi): \mathbb{R}^d \rightarrow \mathbb{R}$ is the joint probability density function.   We consider the eigendecomposition of the above correlation operator as shown below:
\begin{equation}
\int_{0}^T \int_{\Omega} \mathcal{K}(x,t,x',t')\phi_i(x',t')dx' dt'  = \lambda_i \phi_i(x,t), \quad i=1,2, \dots, \infty, \quad  \quad x \in  \Omega,  \quad \mbox{and} \quad t \in [0,T].
\end{equation}
Since $\mathcal{K}$  is a self-adjoint positive operator, its eigenvalues are positive ($\lambda_i=\sigma_i^2 \geq 0$) and its eigenfunctions are orthonormal with respect to the space-time inner product as shown below:
\begin{equation}
  \int_{0}^T \int_{\Omega}  \phi_i(x,t)\phi_j(x,t) dx dt = \delta_{ij}, \quad i,j=1,2, \dots, \infty.
\end{equation}
 Note that in the above definition, the spatial integral is carried out over the entire vasculature. To make this clear, note that the spatial integral can be written as:
 \begin{equation}
 \int_{\Omega} \phi_i(x,t) dx = \sum_{k=1}^K \int_{\Omega^{(k)}} \phi_i(x,t) dx.  
 \end{equation}
The correlation operator $\mathcal{K}$ is global and $x$ and $x'$ could belong to different vessels.  Thus, $\mathcal{K}$ encodes vessel-to-vessel correlations. We choose $\mathcal{K}$ as the customized kernel function for building the regression model for blood flow properties.  In the next section, we present a numerical algorithm to compute the spatiotemporal bases $\phi_i(x,t)$. 

\subsection{Kernel Construction}\label{sec:kernel}
In this section, we present an efficient data-driven methodology to approximate  $\mathcal{K}$. To this end, we express the kernel versus its spectral decomposition as follows:
\begin{equation}\label{eq:kernel}
    \mathcal{K}(x,t,x',t') = \sum_{i=1}^\infty \lambda_i \phi_i(x,t) \phi_i(x',t'),
\end{equation}
where the eigenvalues are sorted in a decreasing order, i.e., $\lambda_1 \geq \lambda_2 \geq \dots$. In the following, we approximate the kernel in a finite-dimensional setting. 
To compute the eigenfunctions $\phi_i(x,t)$, we discretize the spatiotemporal domain. Let $\bm x^{(k)}=[x^{(k)}_1, x^{(k)}_2, \dots, x^{(k)}_n]$  be the vector of discrete points in vessel $(k)$. For simplicity in the notation, we consider the same number of discrete pints in each vessel. However, a different number of points could be considered for each vessel. Therefore, the global discrete spatial vector is simply the union of the local spatial vectors: $\mathbf x = \cup_{i=1}^K  \bm x^{(k)}$, where $\bm x \in \mathbb{R}^{nK\times 1}$.     The temporal domain is also discretized similarly: $\bm t =[t_1,t_2, \dots, t_m]$, where $m$ is the number of time steps. The discrete spatiotemporal domain is obtained via the tensor product of $\bm x$ and $\bm t$, with $N = nK \times m$ points. We reshape the resulting grid into a matrix $\bm X =[ \underline{\bm x} |  \underline{ \bm t}]$, $\underline{\bm x} \in \mathbb{R}^{N \times 1}$,  $\underline{\bm t} \in \mathbb{R}^{N \times 1}$, and $\bm X \in \mathbb{R}^{N \times 2}$.

In the discrete form, each basis function $\phi_i(x,t)$ is represented by a vector $\bs \phi_i\in \mathbb{R}^{N\times 1}$, which contains the values of the basis function on the spatiotemporal grid. Similarly, the spatiotemporal solution of the physics-based model can be represented in the discrete form with vectors of size $\bm u_i \in \mathbb{R}^{N \times 1}$, where $\bm u_i = u(\underline{\bm x},\underline{\bm t};\bs \xi^{(i)})$, where $\bs \xi^{(i)}$ is a random realization of $\bs \xi$. To estimate the expectation operator, we use Monte Carlo sampling:
\begin{equation}
    \mathbb{E}[u(\underline{\bm x},\underline{\bm t},\bs \xi)] \approx \frac{1}{s}\sum_{i=1}^s u(\underline{\bm x},\underline{\bm t},\bs \xi^{(i)}) = \frac{1}{s}\bm U \bm e, 
\end{equation}
where $s$ is the number of Monte Carlo samples, $\{\bs \xi^{(i)} \}_{i=1}^s$ are the Monte Carlo samples of the random vector $\bs \xi$, $\bm U=[\bm u_1, \bm u_2, \dots, \bm u_s] \in  \mathbb{R}^{N \times s}$, and $\bm e=[1,1, \dots, 1]^T\in \mathbb{R}^{N\times 1}$. Therefore, the kernel in the discrete is obtained by applying the above estimators for the expectation operator in Eq. \ref{eq:kernel}, which results in:
\begin{equation}
   \bm K = \frac{1}{s} \bm U \bm U^T,
\end{equation}
where $\bm K \in \mathbb{R}^{N \times N}$.  Typically, $n \sim \mathcal{O}(10^2)$, $m \sim \mathcal{O}(10^2)$ and $K \sim \mathcal{O}(10)$. As a result, $N \sim \mathcal{O}(10^5)$. Consequently, forming the kernel $\bm K$ and computing its eigendecomposition, which is $\mathcal{O}(N^3)$ are cost prohibitive.  Instead, we use the standard procedure by which the eigenvectors of $\bm K$  are computed without forming $\bm K$ explicitly via the singular value decomposition (SVD) of the matrix $\bm U$. As we show in our demonstration examples, the singular values decay quickly for blood flow simulations and as a result, the above decomposition can be truncated at $r < s$ basis functions 
\begin{equation}
\label{DataMatrix}
    \bm U \approx \bs \Phi \bs \Sigma \bm Y^T,
\end{equation}
where $\bs \Phi = [\bs \phi_1, \bs \phi_2, \dots, \bs \phi_r] \in \mathbb{R}^{N\times r}$ are the orthonormal left singular vectors,  $\bs \Sigma \in \mathbb{R}^{r \times r}$ is the matrix of singular values, and $\bm Y \in \mathbb{R}^{s \times r}$ is the matrix of right singular vectors.  The truncation rank ($r$) is determined based on the number of singular values required to approximate matrix $\bm K$ up to a desired accuracy.  Using the above low-rank approximation of $\bm U$, the kernel can be approximated as:
\begin{equation}
\label{Kernel}
   \hat{\bm K} = \frac{1}{s} \bs \Phi \bs \Sigma \bm Y^T \bm Y \bs \Sigma \bs \Phi^T = \bs \Phi \bs \Lambda \bs \Phi^T, 
\end{equation}
where $ \hat{\bm K}$ is a rank-$r$ approximation of $\bm K$. In the above, we use the orthonormality of the right singular vectors, i.e., $\bm Y^T \bm Y = \bm I$ and $\bs \Lambda = 1/s \bs \Sigma^2=\mbox{diag}(\lambda_1, \lambda_2, \dots, \lambda_r)$. One of the key advantages of the above formulation is that kernel $\hat{\bm K}$ does not have to be formed nor stored explicitly. Instead, the matrix of the basis functions ($\bs \Phi)$ and the matrix of eigenvalues  ($\bs \Lambda $) are stored. 

 The above kernel is constructed as a discrete set of spatiotemporal points. However, the kernel needs to be evaluated at $(x,t)$ that may not correspond exactly to one of the spatiotemporal points that $\bm K$ is built for.  It is straightforward to construct \emph{continuous} spatiotemporal kernel. To this end, we build an interpolant for the basis vectors $\bs \phi_i$. This can be done by reshaping  $\bs \phi_i \in \mathbb{R}^{nm\times 1}$ into a matrix of size $n \times m$, which we denote with $[\bs \phi_i] \in \mathbb{R}^{n \times m}$. Let $\bs \Psi(x) = [\psi_1(x), \psi_2(x), \dots, \psi_n(x)]$ be global basis functions that are the union of local basis function in each vessel and $\bs \chi(t) = [\chi_1(t), \chi_2(t), \dots, \chi_m(t)]$ be the temporal basis functions. These basis functions are Lagrange interpolants such that $\psi_i(x^{(k)}_j) = \delta_{ij}$ and $\chi_i(t_j) = \delta_{ij}$. Both $\bs \Psi(x)$ and $\bs \chi(t)$ are \emph{quasimatrices}, which are matrices whose columns are continuous and rows are discrete \cite{trefethen2010householder}. The continuous basis functions can be expressed as:
 \begin{equation}
\hat{\phi}_i(x,t) = \bs \Psi(x) [\bs \phi_i] \bs \chi(t)^T.
\end{equation}
In all the cases considered in this paper, we use piece-wise linear interpolants as the basis spatial and temporal basis functions. 
Therefore, the kernel in the continuous form is given by:
\begin{equation}
\label{Kernel-cont}
   \hat{\mathcal{K}}(x,t,x',t') = \sum_{i=1}^r \lambda_i \hat{\phi}_i(x,t)  \hat{\phi}_i(x',t'). 
\end{equation}
Using the above kernel, Equations \ref{mean} and \ref{covariance}  can be used to perform prediction at new unseen locations and times. To this end, let denote $N$ space, time, and velocity labeled data (observational measurements) with $\bm x =[x_1,x_2, \dots, x_N]$, $\bm t =[t_1,t_2, \dots, t_N]$, and $\bm u =[u_1,u_2, \dots, u_N]$, which means that $u_i$ corresponds to measurement at spatiotemporal coordinate of $(x_i,t_i)$ for $i=1, \dots, N$. Then the GP predictions and posterior uncertainties are:
\begin{subequations}
\begin{equation}
\label{mean-lowrank}
    \mu(x,t) = \hat{\mathcal{K}}\left(x,t, \bm x,\bm t \right) (\hat{\mathcal{K}}\left(\bm x,\bm t, \bm x,\bm t \right)+ \sigma^2_n \mathbf{I})^{-1} \mathbf{u},
\end{equation}
\begin{equation}
\label{covariance-lowrank}
    \sigma^2(x,t) =  \hat{\mathcal{K}}\left(x,t,x,t\right)-\hat{\mathcal{K}}\left(x,t, \bm x,\bm t \right) (\hat{\mathcal{K}}\left(\bm x,\bm t, \bm x,\bm t \right)+ \sigma^2_n \mathbf{I})^{-1} \hat{\mathcal{K}}\left(\bm x,\bm t, x,t \right),
\end{equation}
\end{subequations}
where $\hat{\mathcal{K}}\left(\bm x,\bm t, \bm x,\bm t \right) \in \mathbb{R}^{N \times N}$ and $\bm I \in \mathbb{R}^{N \times N}$ is the identity matrix.  The variance of the noise ($\sigma_n^2$) can be determined by maximizing the negative likelihood. The presented methodology addresses all of the issues raised in the introduction (§\ref{sec:Intro}) and offers the following advantages: 

\begin{enumerate}
\item The constructed kernel $\hat{\mathcal{K}}$
  encodes both spatiotemporal and vessel-to-vessel correlations. This capability enables the reconstruction of blood flow velocity across the entire vasculature using a very limited number of measurements. For example, it allows for the estimation of blood flow in vessels that lack direct measurements.\\
   
  \item Any reconstructed flow automatically satisfies the conservation of mass. To demonstrate this, note that the posterior mean from Eq. \ref{mean-lowrank} can be expressed as a linear combination of the spatio-temporal basis functions, written as:
  \begin{equation}\label{eq:mean-lin_bas}
     \mu(x,t) = \hat{\mathcal{K}}\left(x,t, \bm x,\bm t \right) \bm z = \sum_{j=1}^N \sum_{i=1}^r \lambda_i \hat{\phi}_i(x,t) \hat{\phi}_i(x_j,t_j) y_j = \sum_{i=1}^r w_i \hat{\phi}_i(x,t),
  \end{equation}
  where $\bm z =[z_1, \dots, z_N]^T= (\hat{\mathcal{K}}\left(\bm x,\bm t, \bm x,\bm t \right)+ \sigma^2_n \mathbf{I})^{-1} \mathbf{u}$ and $w_i = \lambda_i \sum_{j=1}^N \hat{\phi}_i(x_j,t_j)z_j$. From Eq. \ref{DataMatrix}, the basis functions are themselves a linear combination of the simulated data, i.e., $\hat{\phi}_i(x,t) = \sum_{j=1}^s u_j(x,t)  Y_{ji}/\sigma_i$, where $u_j(x,t) = \bs \Psi(x) [\bm u_j] \bs \chi(t)^T$. Replacing this expression of the basis function into Eq. \ref{eq:mean-lin_bas}, results in:
  \begin{equation}\label{eq:mean-lin_sample}
     \mu(x,t) =  \sum_{j=1}^s a_j u_j(x,t), 
  \end{equation}
  where $a_j = \sum_{i=1}^r w_i Y_{ji}/\sigma_i$.
This shows that the posterior mean can be expressed as a linear combination of all $s$ samples. Given that each sample  $ u_j(x,t)$ satisfies the conservation of mass, their linear combination also preserves the conservation of mass.
  
    \item Since the number of observations is typically small, the computational cost of calculating the posterior means and uncertainty is negligible. The offline costs consist of running 1D simulations and performing SVD on the generated data to create the kernel. For instance, on a system equipped with a 3.4 GHz Quad-Core Intel Core i5 processor, running each 1D simulation of an abdominal aorta vasculature network, which consists of 17 vessels, takes 28 seconds. A total of 150 simulations can be run in parallel. Creating the kernel requires 41 seconds. The simulation time for each sample scales roughly linearly with the number of vessels. The demonstration cases presented in this paper involve a relatively small number of vessels (less than 30). However, for vasculature networks with a larger number of vessels, recently developed low-rank approximation methods based on time-dependent bases may be employed to accelerate the simulation time \cite{donello2023oblique}.
    

\end{enumerate}

\subsection{Uncertainty Characterization}
\label{sec:UQ}
To generate the kernel as defined in Eq.~\ref{Kernel}, we calculate the blood flow properties using Eqs.~\ref{conMass1d}, \ref{Consmoment1d}, and \ref{press1d} for \(s\) number of samples. Below, we detail how the stochastic modeling accounts for the uncertainty in precise inlet and outlet boundary conditions.

 The human arterial system consists of many vessels and in many cases, it is either impossible or unnecessary to simulate them all. The well-known three-element Windkessel model (RCR), allows us to simulate a part of the vasculature network and apply the effect of neglected vessels at the terminals. In this model, if we denote the total resistance by $R_t=R_1 + R_2$ and compliance by $C$, for the outlet:
\begin{equation}
    p + R_2C \frac{dp}{dt} - R_tQ - p_{\infty} - R_1R_2C \frac{dQ}{dt}=0.
    \label{BC}
\end{equation}
Here, $Q=Au$ is the flux at the outlet, $p_{\infty}$ is the downstream pressure, and $R_1= \frac{\rho c_0}{A_0}$ where $c_0=\sqrt{\frac{\beta}{2\rho \sqrt{A_0}}}$ is the speed of wave propagation. In the Nektar solver, this model is adapted to the characteristic waves at the terminals \cite{alastruey2008reduced}.


We randomize the inlet velocity, outflow boundary conditions ($R_t$ and $C$), and cross-section areas. For example Figure \ref{Fig:Inputvelex1} shows the range of randomized inlet velocity for a Y-shaped vessel shown in Figure \ref{Fig:Geometryex1}. In this figure, the periodic velocity with the periodic time $T$ is the summation of $n_i$ component functions, each described as follows:
\begin{equation}
 u(t; \bs \xi)=  a_0 + \sum_{i=1}^{n_i} a_i \exp (-\frac{(t-[t/T]T- b_i)^2}{c_i}),
 \label{inlet}
\end{equation}
where $[z]$ is the rounded integer of $z$ such that $ 0 \leq z-[z]<1$,   $a_i$, $b_i$, and $c_i$ adjust peaks/valleys, their location, and their sharpness. We randomize the inlet profile by randomizing the parameters $[a_0,a_i,b_i,c_i]$ as shown below:
\begin{equation}
    \xi_k= \overline{\xi}_k +\sigma_k \psi, \quad \mbox{where} \quad \xi_k \equiv [a_0,a_i,b_i,c_i], \quad i=1, \dots, n_i.
    \label{abcrand}
\end{equation}
Here $\psi=\mathcal{U}[-0.5,0.5]$ is a zero-mean uniform random variable with unit variance and $\sigma_k$ is the standard deviation of variable $\sigma_k \psi$,  $\xi_k$ is the $k$th random variable and $\overline{\xi}_k = \mathbb{E}[\xi_k]$ is the mean of $\xi_k$. Therefore, the inlet randomization introduces $k=1, \dots 3n_i+1$ random variables.  We randomize the outflow parameters $(R_t,C)$ for the vessels  that have outflow boundary conditions as follows: 
\begin{equation}
    \xi_k= \overline{\xi}_k (1 +  \sigma_k \psi) , \quad \mbox{where} \quad \xi_k \equiv [R_t,C]. 
    \label{area_outlet}
\end{equation}
Therefore, each vessel that has an outflow boundary condition introduces two more random variables that will be appended to the existing random variables.  We also randomize the initial area $\xi_k = \{ A_{0_i}\}, i=1,2, \dots, K$ for each vessel in a form similar to Eq. \ref{area_outlet}. This will also add $K$ (number of vessels) random variables.   Using these parameters, we perform 1D simulations and store the resulting data in the matrix $\mathbf{U}$ (Eq. \ref{DataMatrix}) and follow the algorithm presented in Section \ref{sec:kernel} to construct kernel $\hat{\mathcal{K}}$.

\section{Demonstrations}\label{sec:problem}
In this section, we demonstrate the performance of the presented methodology in three cases: i) a Y-shaped vessel, where measurements are taken from 1D simulations; ii) the lower part of the abdominal aorta vasculature, where measurements are taken from 3D simulation; and iii) the brain vasculature (Circle of Willis), where measurements are taken from 4D Flow MRI. In the following examples, we refer to the set of data points in the form of $(\bm{x}, \bm{t}, \bm{u})$, used in Eqs. \ref{mean-lowrank}-\ref{covariance-lowrank}, as \emph{measurement}, and we refer to points at which we perform comparisons as \emph{prediction}. In all cases, the kernel is constructed using stochastic 1D simulations. Once the kernel is constructed, we obtain measurement data, i.e., $(\bm{x}, \bm{t}, \bm{u})$, from various sources, including 1D simulation, 3D simulation, and 4D Flow MRI and we perform predictions using Eqs. \ref{mean-lowrank}-\ref{covariance-lowrank}. 
\subsection{Y-Shaped Vessel}
\begin{figure}[t!]
\centering
\subfigure[Y-shaped bifurcation schematic]{
  \raisebox{0.1cm} {\includegraphics[clip=true ,width=0.31\textwidth]{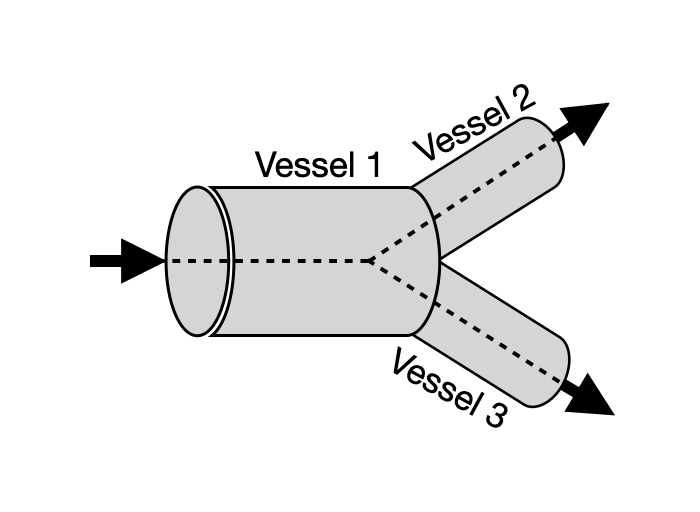}}
\label{Fig:Geometryex1}
}
\subfigure[Inlet velocity]
{
\includegraphics[clip=true, width=0.31\textwidth]{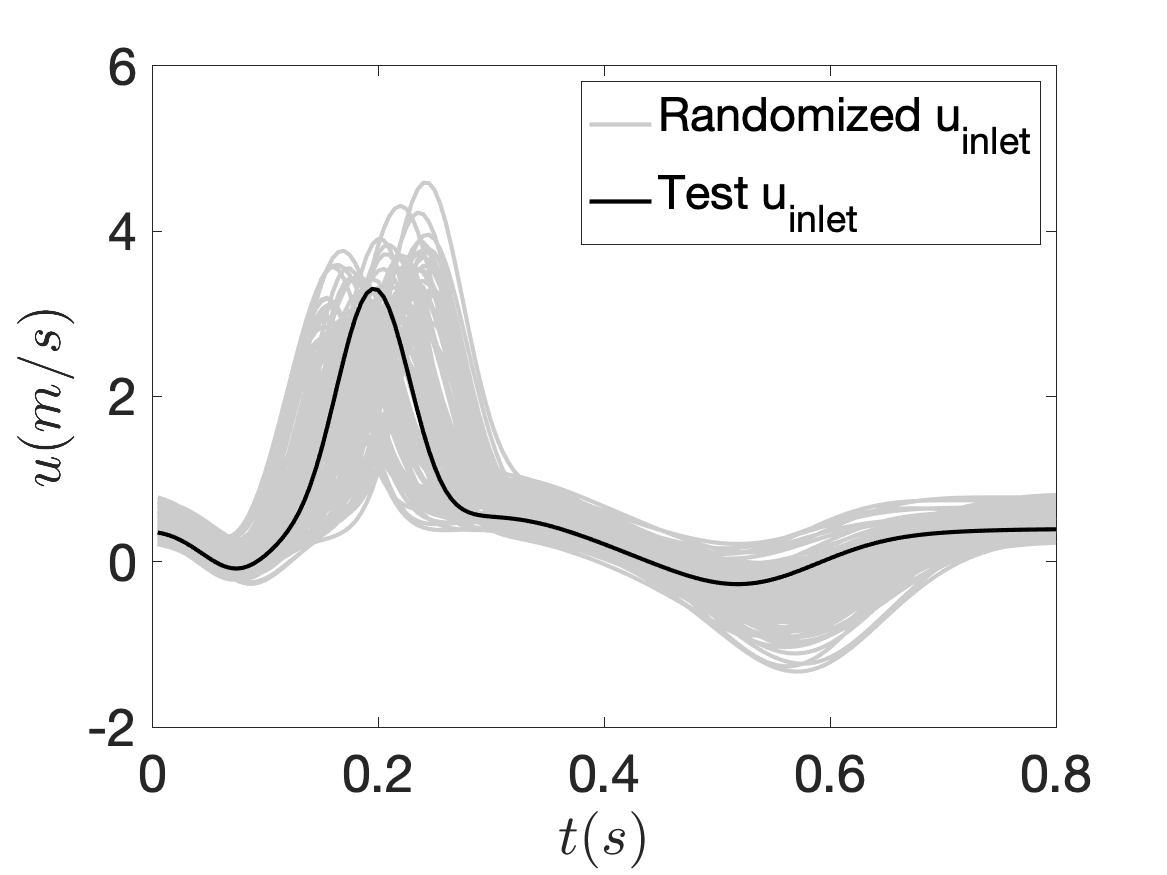}
\label{Fig:Inputvelex1}
}
\subfigure[Singular values]{
\includegraphics[width=0.31\textwidth]{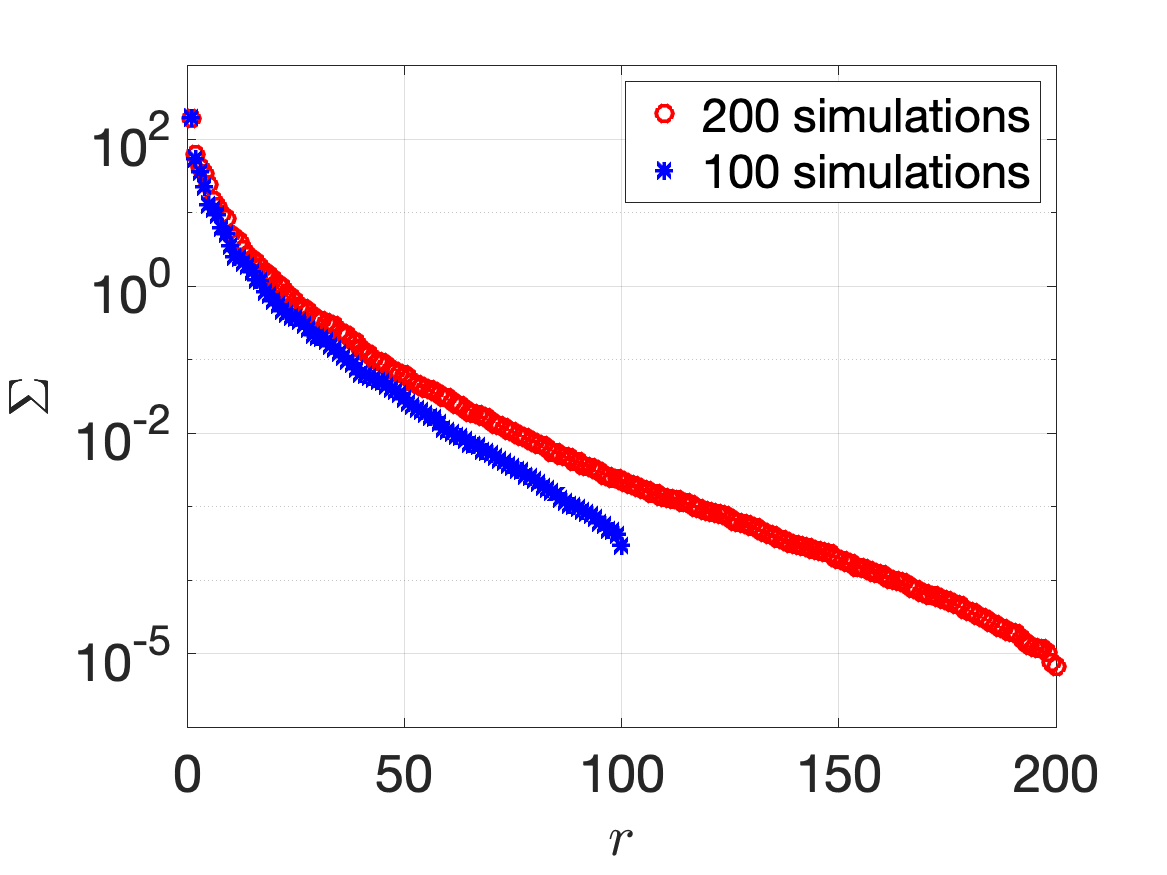}
\label{Fig:SV_vs_sim}
}

\caption{a) Y-shaped vessel schematic. b) Y-shaped vessel random inlet velocities for all samples and the randomly selected sample as the measurement for prediction and validation. c)  The dominant singular values remain unchanged across different numbers of simulations.}
\label{Fig:example1setup}
\end{figure}

\begin{figure}[t!]
\centering
\subfigure[Case 1]{
\includegraphics[clip=true ,width=0.3\textwidth]{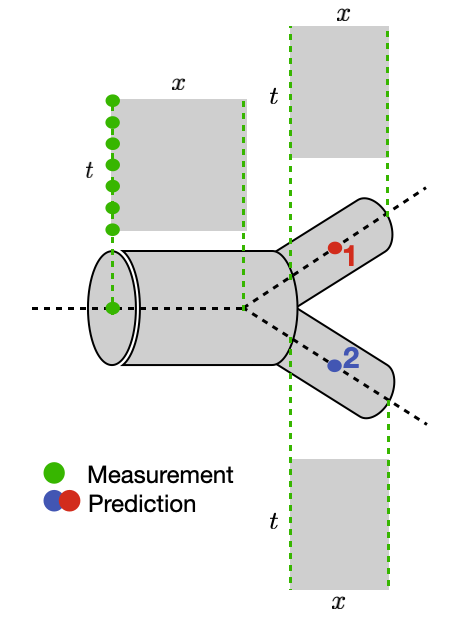}
\label{Fig:case1}
}
\subfigure[Case 1, point 1]{
  \raisebox{1cm} {\includegraphics[trim=2cm 0cm 2cm 0cm,clip=true ,width=0.3\textwidth]{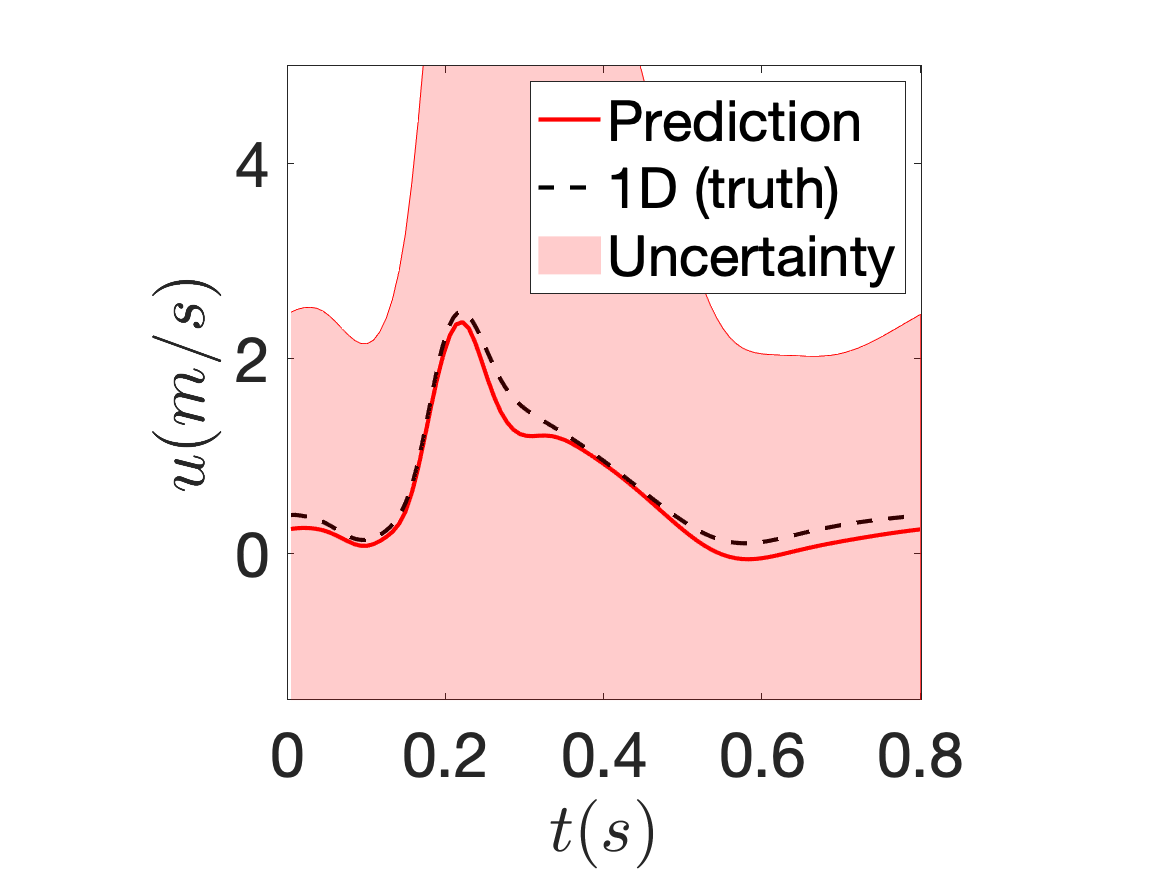}}
\label{Fig:case1V2}
}
\subfigure[Case 1, point 2]{
\raisebox{1cm} {\includegraphics[trim=2cm 0cm 2cm 0cm, clip=true ,width=0.3\textwidth]{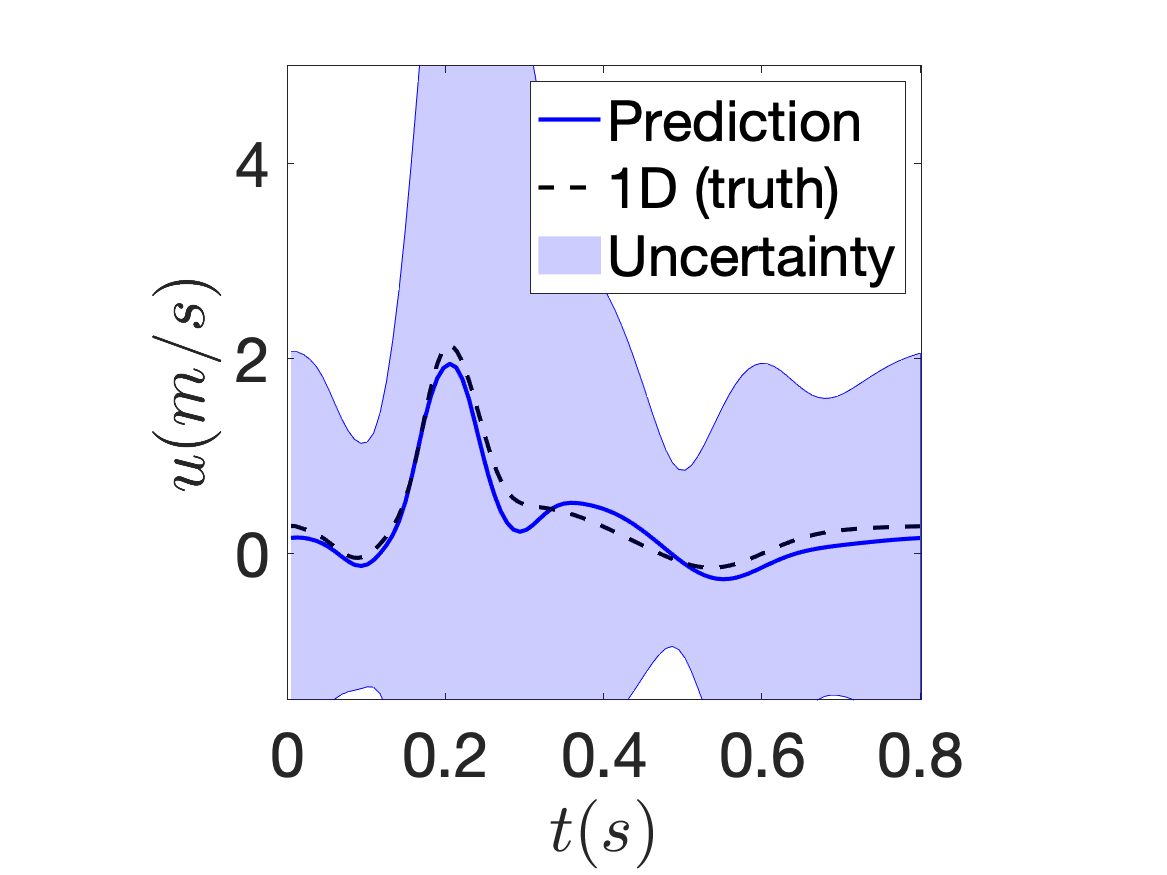}}
\label{Fig:case1V3}
}

\subfigure[Case 2]{
\includegraphics[clip=true ,width=0.3\textwidth]{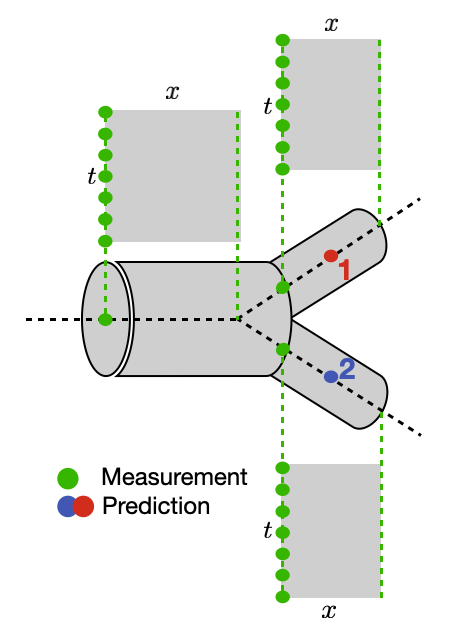}
\label{Fig:case3}
}
\subfigure[Case 2, point 1]{
\raisebox{1cm} {\includegraphics[trim=2cm 0cm 2cm 0cm, clip=true ,width=0.3\textwidth]{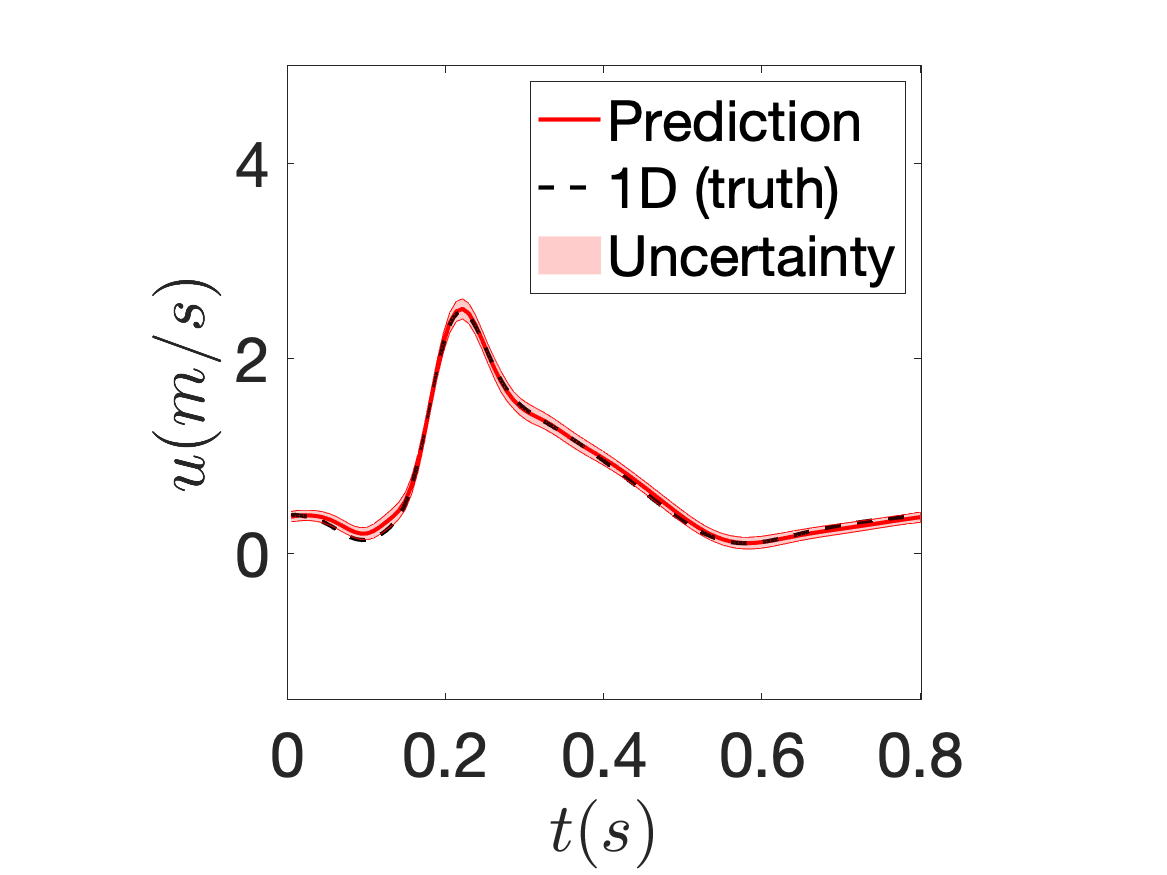}}
\label{Fig:case3V2}
}
\subfigure[Case 2, point 2]{
\raisebox{1cm} {\includegraphics[trim=2cm 0cm 2cm 0cm, clip=true ,width=0.3\textwidth]{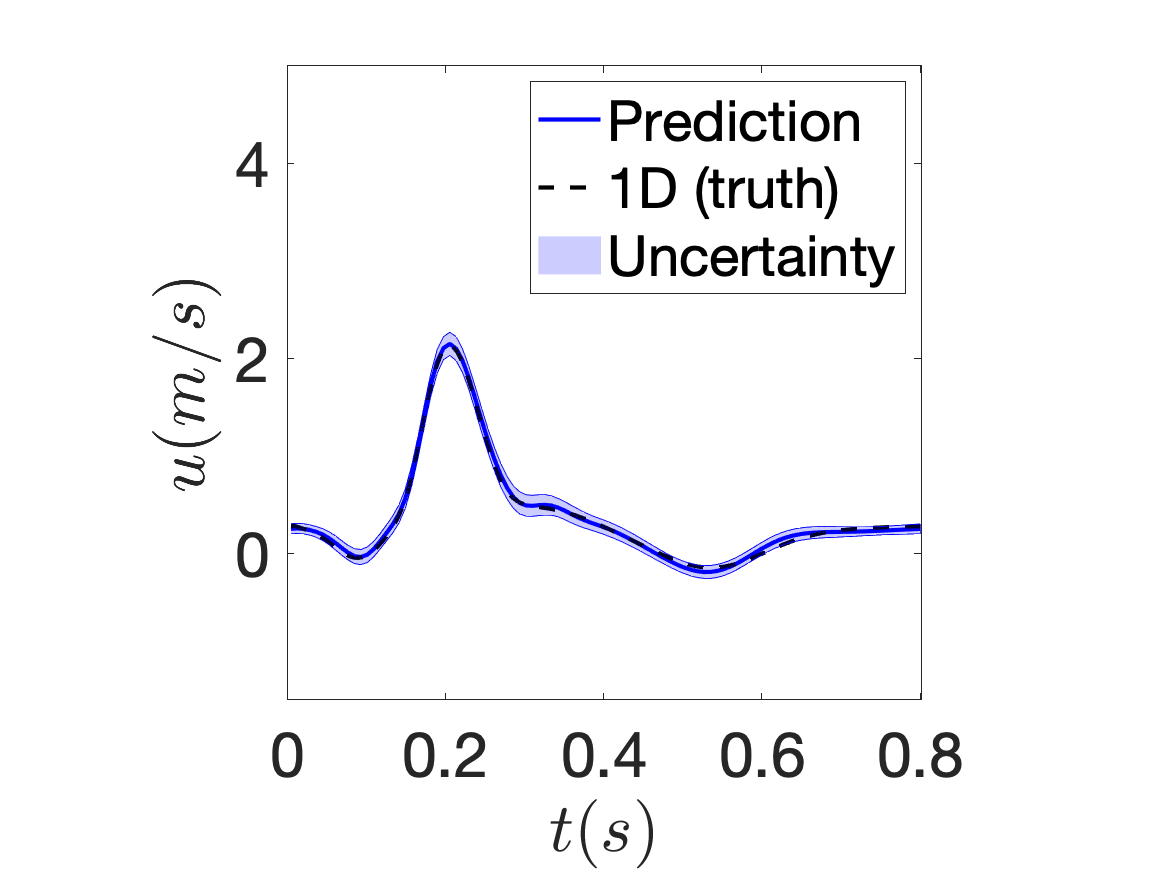}}
\label{Fig:case3V3}
}

\caption{Schematic of the measurement positions. a) Case 1 with high temporal and low spatial resolution. The methodology provides predictions for all spatiotemporal locations, and we have chosen two specific prediction points for result comparison.  b and c) Compare the predicted velocity at prediction points for Case 1 which has the most uncertainty due to the lack of nearby measurements. d) Case 2 with low spatiotemporal resolution.  e and f) Compare the predicted velocity at prediction points for Case 2 which is improved compared with Case 1.}
\label{Fig:Resultex1}
\end{figure}

In the first test case, we consider a Y-shaped vessel. The schematic of the problem is shown in Figure \ref{Fig:Geometryex1}, which is similar to the first example in \cite{kissas2020machine}. The equilibrium area, length, and Windkessel parameters are presented in Table \ref{table:Geo1}. To create the kernel, we generate data using the 1D model (Eqs.~\ref{conMass1d}-\ref{Consmoment1d}), wherein the inlet velocity, cross-sectional areas, resistance, and compliance are randomized using Eqs. \ref{abcrand} and \ref{area_outlet}. For the inlet velocity (Figure \ref{Fig:Inputvelex1}), $T=0.8s$, $n_i=4$, and its randomized parameters values are listed in Table \ref{table:rand1} along with $R_t$, $C$, and $A_0$. We generate $s=200$ samples and collect their simulation results. The results are stored for $n=100$ equidistant points in each vessel and $m=160$ equidistant time snapshots to create the kernel based on Eqs. \ref{DataMatrix} and \ref{Kernel}. The rank is determined as the smallest $r$ for which $\sum_{i=1}^r\sigma^2_i/\sum_{i=1}^s\sigma^2_i > 0.99$. 

To perform prediction, we randomly select a new sample of the vector $\bs \xi$ and we denote this sample with $\bs \xi^*$. We perform the 1D simulation for $\bs \xi = \bs \xi^*$. We use the spatiotemporal result of this sample as the ground truth. As we explain below, we use only a small number of spatiotemporal data points generated by this simulation as the input data to the GP model.  In particular, using the kernel created and a few measurements from the selected 1D simulation data, we perform predictions for all three vessels. A \(5\%\) Gaussian noise is added to the measurements to intentionally corrupt them with noise, mimicking real data. 
 Figure \ref{Fig:SV_vs_sim} compares the singular values from two different numbers of simulations: $s=200$ (our case) and $s=100$. Based on this figure, the leading singular values of the two cases are very close to each other, implying that the number of samples is sufficient.

The GP model generates continuous spatiotemporal velocity predictions throughout the entire network. For comparison, we present the GP prediction results for only two representative points. Specifically, we compare the GP predictions at points 1 and 2, located in the middle of each corresponding vessel, as illustrated in Figure~\ref{Fig:case1} and Figure~\ref{Fig:case3} for the following two scenarios:

\begin{itemize}
    \item Case 1 (low spatial resolution and high temporal resolution): The schematic of the spatiotemporal coordinates of the measurement points is illustrated in the $x-t$ coordinate system in Figure \ref{Fig:case1}. In particular, we utilize the inlet data from vessel 1 with a temporal resolution of $\Delta t = 0.005s$. The primary aim of this case is to address scenarios involving measurements with very low spatial resolution, in this instance, the extreme case of having only the time series data of one spatial point.   According to Figures \ref{Fig:case1V2} and \ref{Fig:case1V3}, the predicted velocity shows good agreement with the truth, although it exhibits significant uncertainty due to the insufficiency of measurements. It is noteworthy that in this case, despite no measurements being used in vessels 2 and 3, the GP model accurately predicts the velocity in those vessels. This is attributed to the fact that the kernel is constructed globally which encodes the vessel-to-vessel correlations.

    
    \item Case 2 (low spatio-temporal resolution): In this scenario, we analyze a situation with low spatio-temporal resolution measurements, though not as extreme as Case 1 (Figure \ref{Fig:case3}). Specifically, we utilize the inlets of all vessels for measurements with $\Delta t=0.01s$. As illustrated in Figures \ref{Fig:case3V2} and \ref{Fig:case3V3}, predictions at points 1 and 2 are very close to the ground truth, and the uncertainties are significantly reduced since more measurements are used in comparison to Case 1.

\end{itemize}


\begin{table}[t!]
\caption{Y-shaped vessel 1D simulation specification.}
\centering
\begin{tabular}{| c | c | c | c | c | c |}
\hline
Vessel & Length$(m)$  & $A_0$ $(m^2)$& Compliance$(10^{-10}\frac{m^3}{Pa})$ & Total resistance$(10^{10}\frac{Pa.s}{m^3})$ & $\beta(\frac{Pa}{m})$ \\ [0.5ex] 
\hline
\rule{0pt}{4ex} 
1&0.1703& $A_{0_1}=$ 1.36e-5 & - & - & 6.97e7\\
2&0.007& $A_{0_2}=$1.81e-6 & $C_1=$0.3428 & $R_{t_1}=$1.19  & 5.42e8\\
3&0.00667& $A_{0_3}=$1.36e-5 & $C_2=$0.6661 & $R_{t_2}=$0.2702 & 6.96e7\\
\hline
\end{tabular}
\label{table:Geo1}
\end{table}

\begin{table}[t!]
\caption{Randomized parameters for the Y-shaped vessel network. }
\centering
\resizebox{\textwidth}{!}{%
\begin{tabular}{| c | c | c | c | c | c | c | c |}
\hline
k & $a_0$ & $a_i$ & $b_i$ & $c_i$ & $R_{t_i} $ & $C_i $ & $A_{0_i}$\\ [0.5ex] 
\hline
\rule{0pt}{4ex} 
$\overline{\xi}_k$ & 0.5 &  [-0.5,3,-1,-0.1]& [0.08,0.2,0.4,0.6] & [2e-3,5e-3,1.5e-2,0.01] & [$R_{t_1}$,$R_{t_2}$] & [$C_1$,$C_2$] & [$A_{0_1}$,$A_{0_2}$,$A_{0_3}$]\\

$\sigma_k$& 0.5 & [0,0.9,0.5,0.9] & [0.02,0.1,0.15,0.3] & [0,1e-3,1e-3,0] & [0.05,0.05] & [0.05,0.05] & [0.5,0.5,0.5]\\
\hline
\end{tabular}
}
\label{table:rand1}
\end{table}

\subsection{Abdominal Aorta}
\begin{figure}[t!]
\centering
\subfigure[Schematic]{
\includegraphics[width=0.3\textwidth]{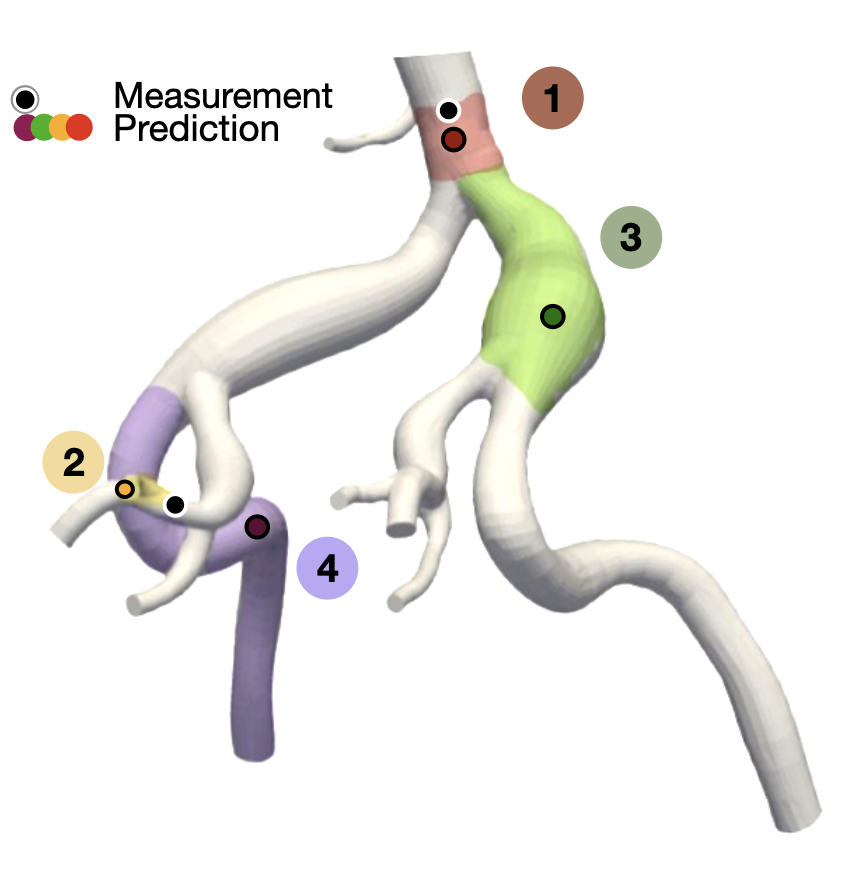}
\label{Fig:Geometryex2}
}
\subfigure[1D map]{
\includegraphics[ clip=true, width=0.3\textwidth]{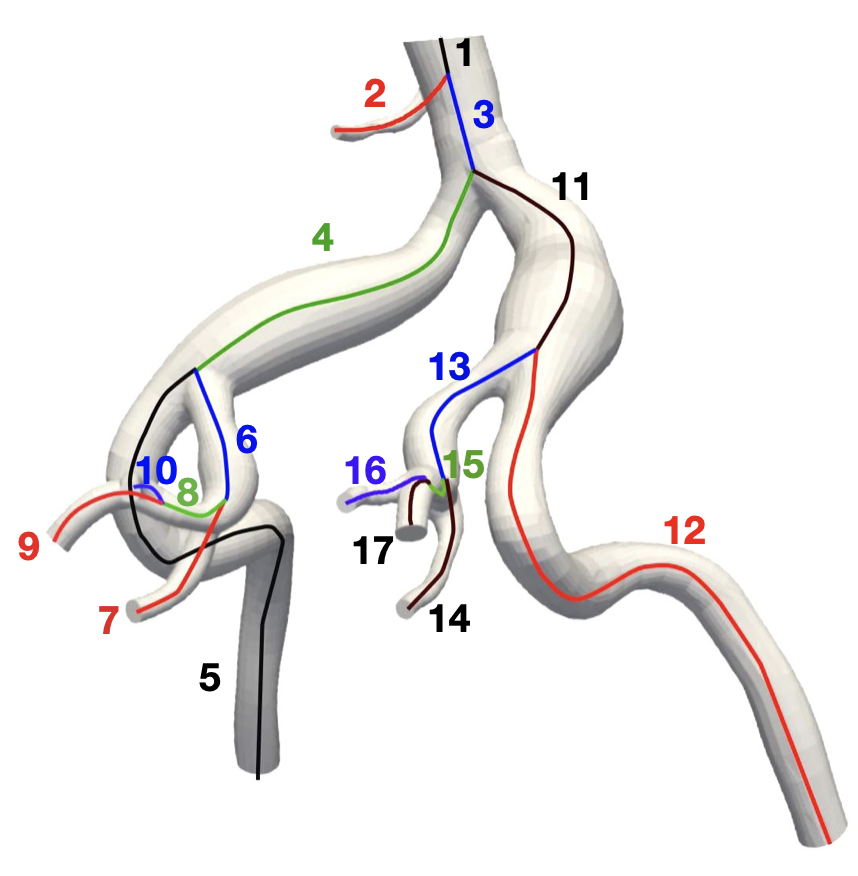}
\label{Fig:AAmap}
}
\subfigure[Inlet velocity]{
\raisebox{0.75cm} {\includegraphics[ clip=true, width=0.3\textwidth]{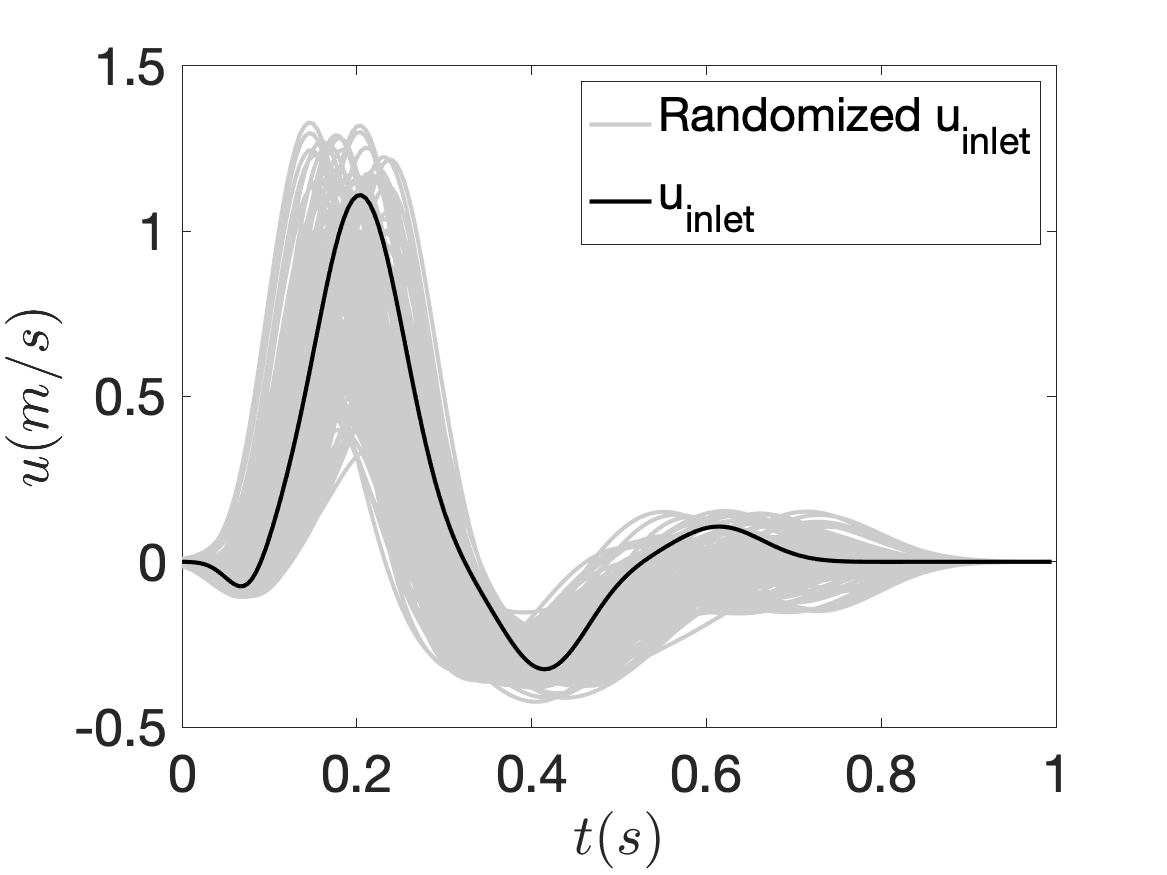}}
\label{Fig:inletex2}
}
\caption{a) Abdominal aorta schematic with the location of the velocity measurements and prediction. b) 1D map of the abdominal aorta, extracted from the vessel's centerline. c) Random inlet velocities and the equivalent inlet velocity for 3D simulation.}
\label{Fig:example2setup}
\end{figure}

\begin{table}[t!]
\caption{Randomized parameters for abdominal aorta.}
\centering
\resizebox{\textwidth}{!}{%
\begin{tabular}{| c | c | c | c | c | c | c | c |}
\hline
k & $a_0$ & $a_i$ & $b_i$ & $c_i$ & $R_{t_i} $ & $C_i$ & $A_{0_i}$\\ [0.5ex] 
\hline
\rule{0pt}{4ex} 
$\overline{\xi}_k$ & 0 &  [-0.1,0.9,-0.3,0.01]& [0.08,0.2,0.4,0.6] & [2e-3,5e-3,1.5e-2,0.01] & [$R_{t_1}$,...,$R_{t_9}$] & [$C_1$,...,$C_9$] & [$A_{0_1}$,...,$A_{0_{17}}$]\\

$\sigma_k$& 1e-3 & [0,0.9,0.5,0.9] & [0.02,0.1,0.15,0.3] & [0,1e-3,1e-3,0] & [0,...,0] & [0,...,0] & [0,...,0]\\
\hline
\end{tabular}
}
\label{table:rand2}
\end{table}

\begin{figure}[t!]
\centering
\subfigure[Point 1, 3D]{
\includegraphics[ clip=true ,width=0.22\textwidth]{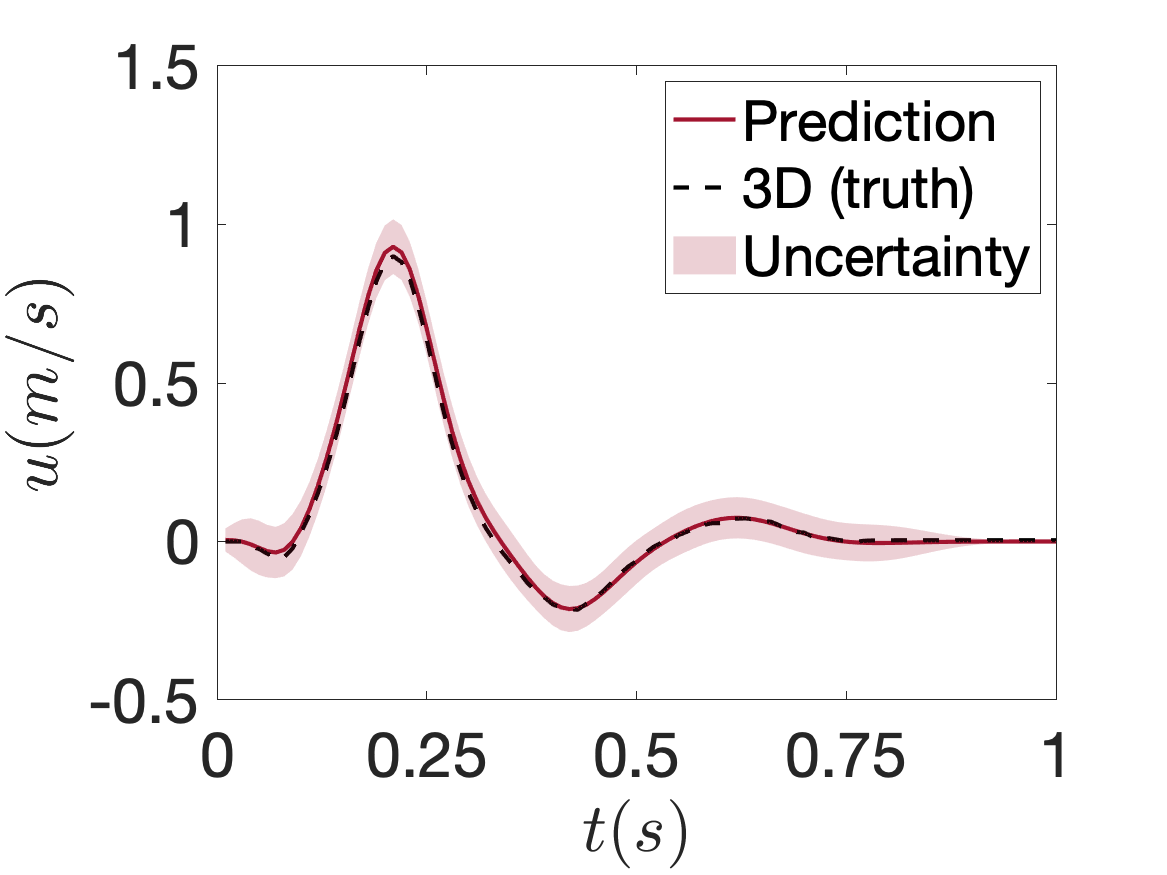}
\label{Fig:Abd3U}
}
\subfigure[Point 2, 3D]{
\includegraphics[ clip=true ,width=0.22\textwidth]{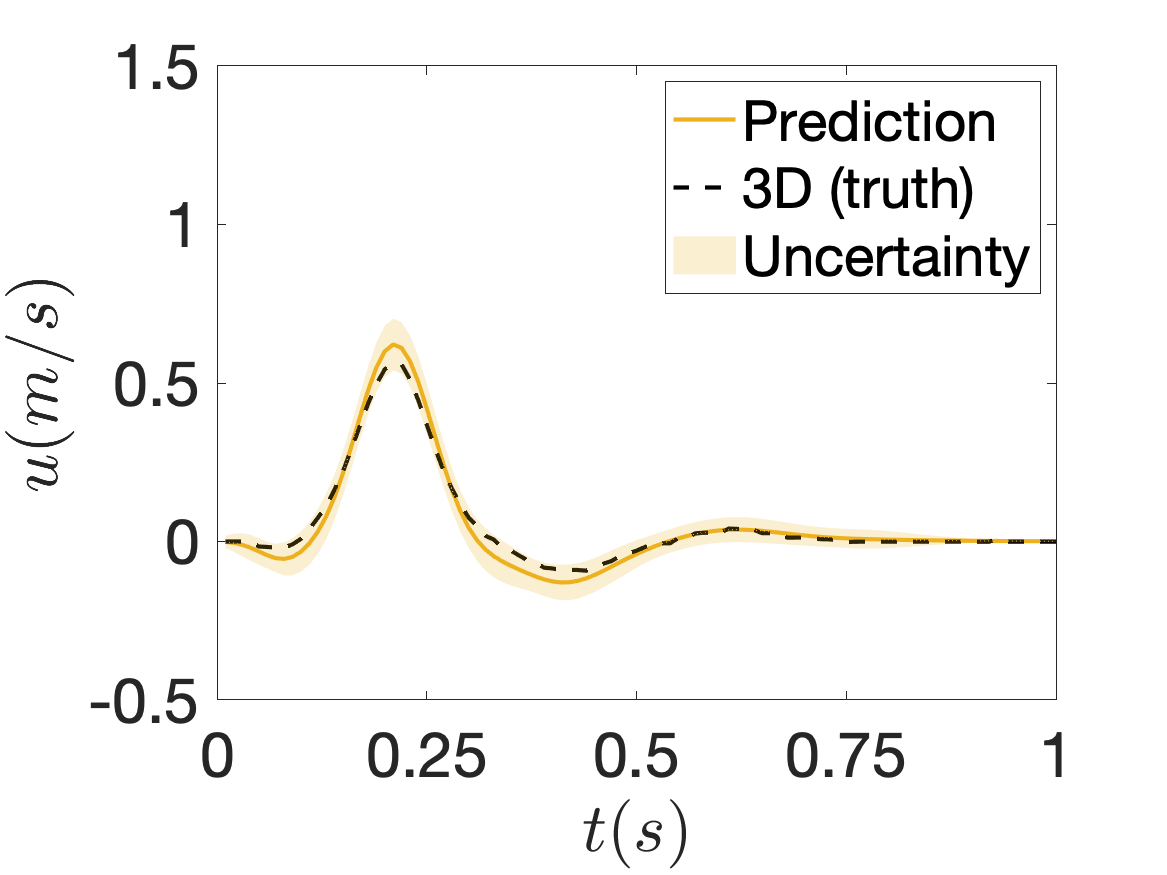}
\label{Fig:Abd10U}
}
\subfigure[Point 3, 3D]{
\includegraphics[ clip=true ,width=0.22\textwidth]{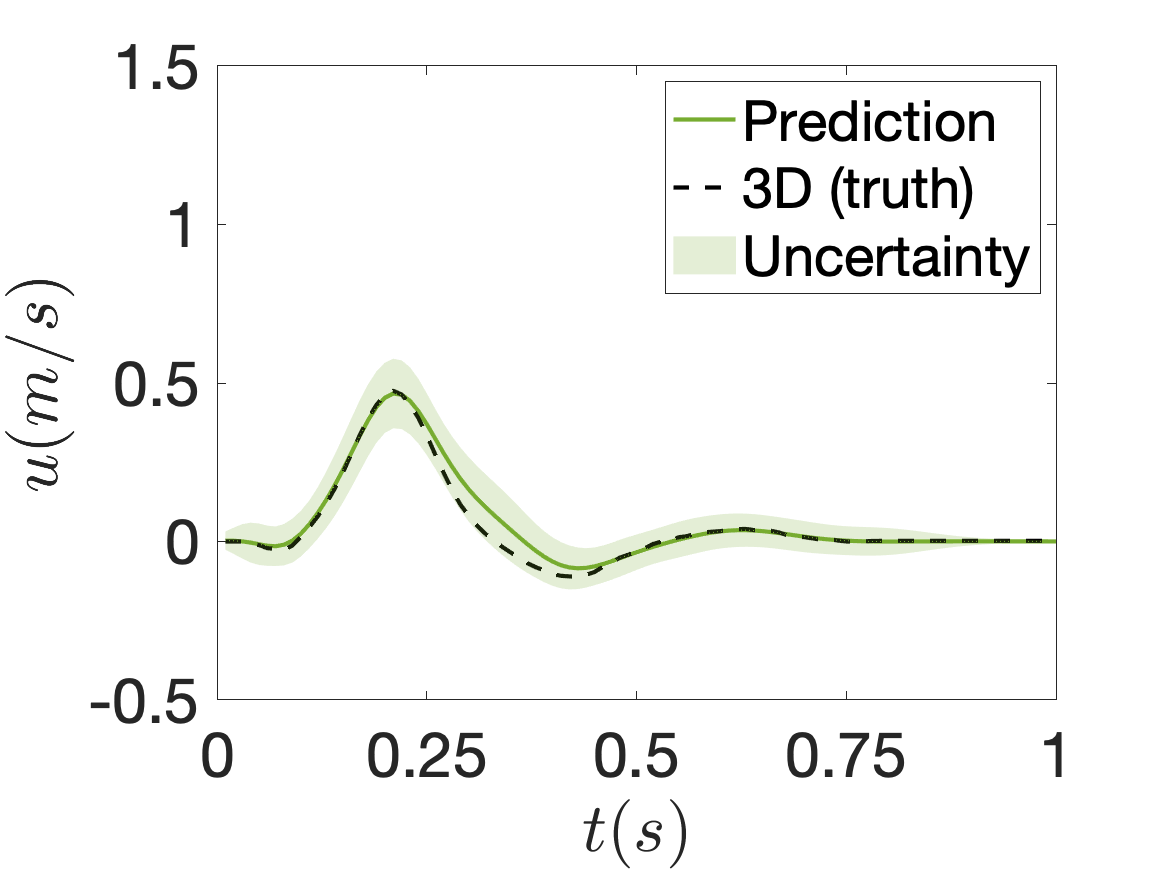}
\label{Fig:Abd11U}
}
\subfigure[Point 4, 3D]{
\includegraphics[ clip=true ,width=0.22\textwidth]{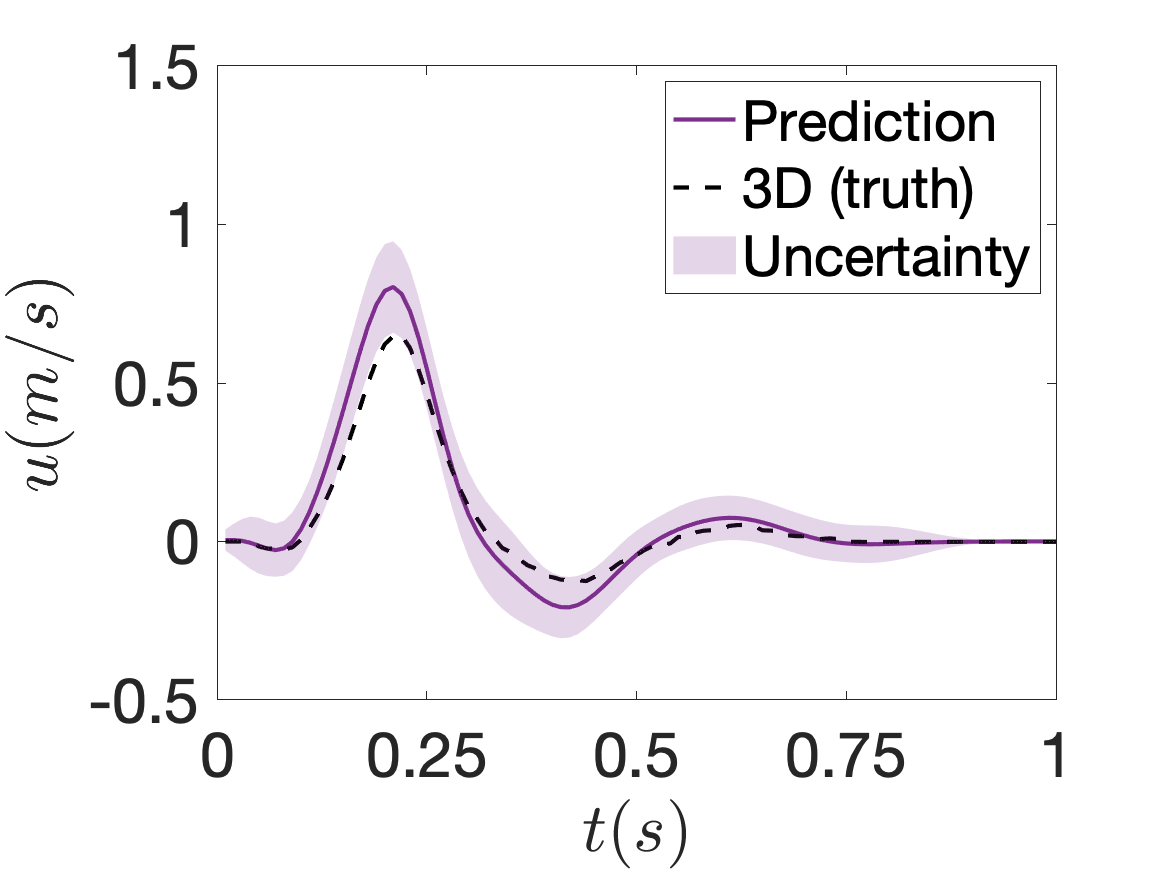}
\label{Fig:Abd5U}
}
\caption{Case 1: abdominal aorta prediction using 3D simulation as measurement. Predictions are compared with 3D simulation as the ground truth at a) point 1, b) point 2, c) point 3, and d) point 4. The location of these points and utilized measurements are indicated in Figure \ref{Fig:Geometryex2}. In these comparisons, despite the kernel relying on 1D simulations the GP predictions align with 3D simulations across various velocity ranges.}
\label{Fig:Abdresult_3D}
\end{figure}

\begin{figure}[t!]
\centering
\subfigure[Point 1, 1D]{
\includegraphics[  clip=true ,width=0.22\textwidth]{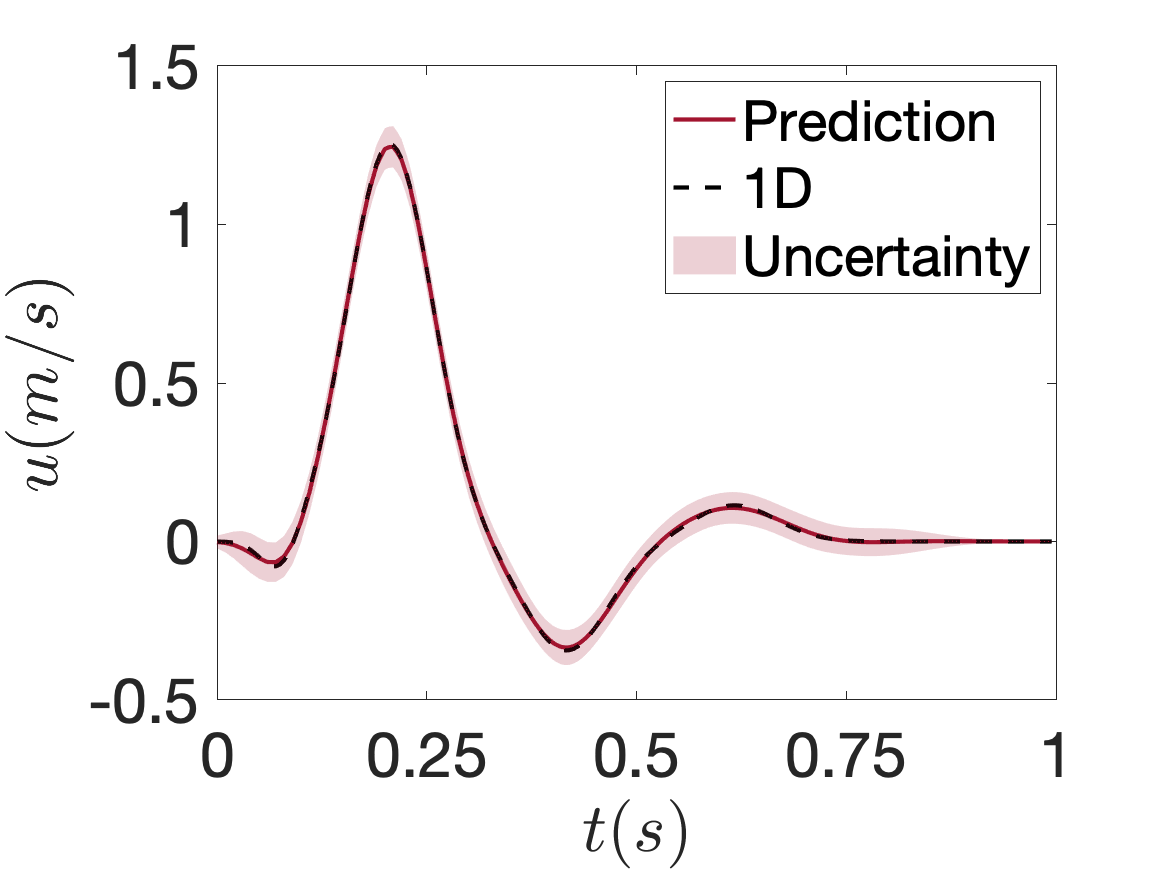}
\label{Fig:Abd3U1d}
}
\subfigure[Point 2, 1D]{
\includegraphics[clip=true ,width=0.22\textwidth]{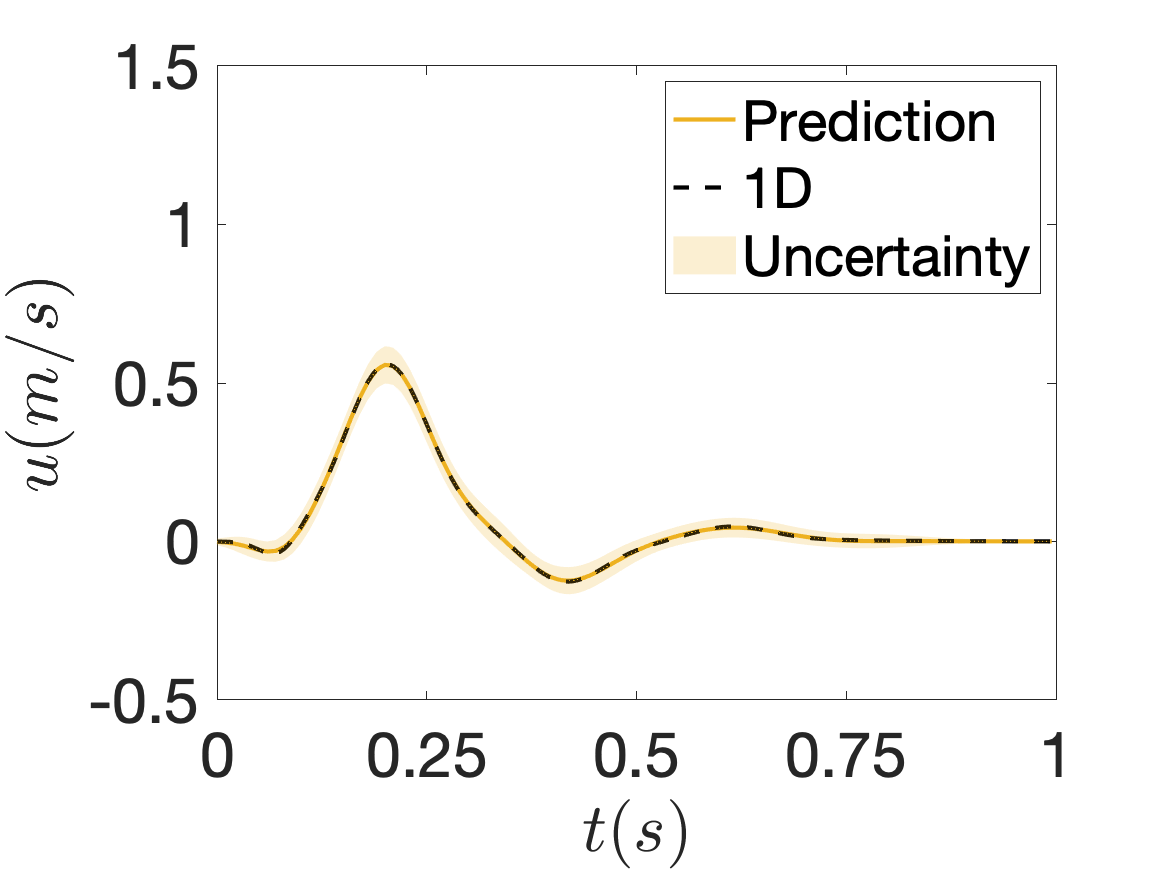}
\label{Fig:Abd10U1d}
}
\subfigure[Point 3, 1D]{
\includegraphics[clip=true ,width=0.22\textwidth]{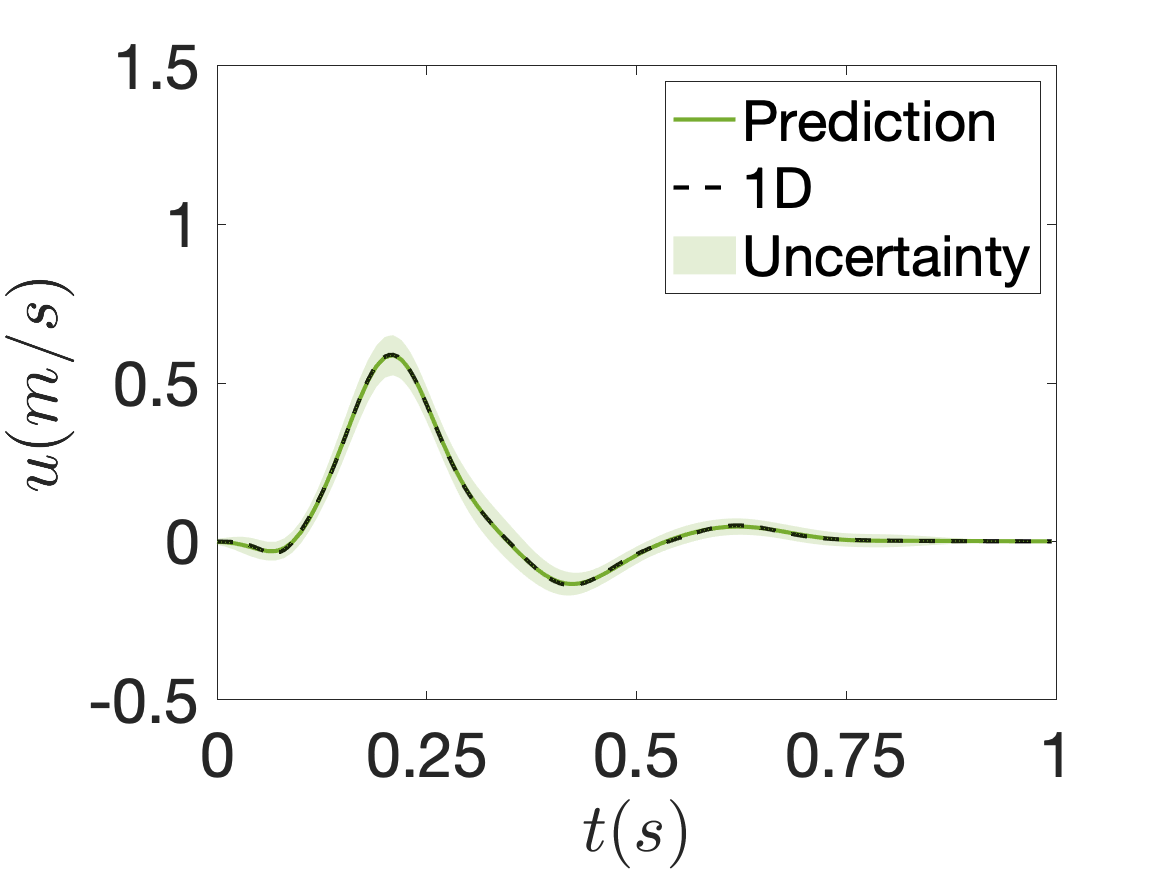}
\label{Fig:Abd11U1d}
}
\subfigure[Point 4, 1D]{
\includegraphics[clip=true ,width=0.22\textwidth]{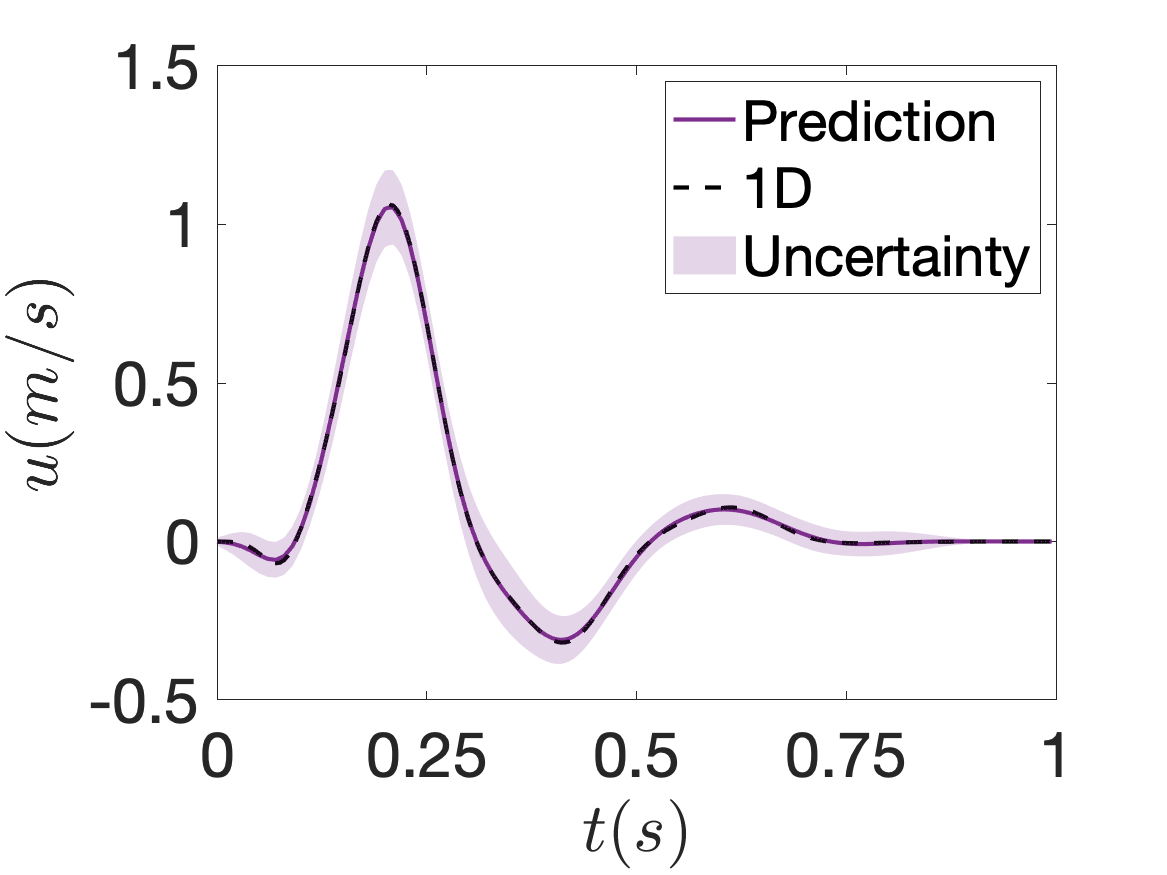}
\label{Fig:Abd5U1d}
}
\caption{Case 2: abdominal aorta prediction using 1D simulation as measurement. The location of these points and utilized measurements are the same as Case 1 and is indicated in Figure \ref{Fig:Geometryex2}. Comparisons are made using 1D simulation as the ground truth at a) point 1, b) point 2, c) point 3, and d) point 4. In these comparisons the GP predictions matches with 1D simulations.}
\label{Fig:Abdresult_1d}
\end{figure}

In this case, we predict blood flow in the abdominal aorta and iliac branches, where the measurements (ground truth) are derived from 3D simulation. To extract the geometry of the problem (Figure~\ref{Fig:Geometryex2}), we utilize time of flight (ToF) MRI data from the aortofemoral vasculature of a 67-year-old male subject. This data spans from the thoracic aorta to the femoral artery bifurcation and includes the thoracic, abdominal, renal, and femoral arteries (the model is available at www.vascularmodel.com). The voxel resolution in the right-left, anterior-posterior, and superior-inferior directions is $0.78mm$, $2.00mm$, and $0.78mm$, respectively. 3D simulation is conducted using direct numerical simulation (DNS) of the 3D Navier-Stokes equations, assuming the flow is Newtonian and incompressible, with the spectral/hp element method. Specifically, we employ tetrahedral elements with a polynomial order of $p=3$. The generated mesh comprises $Ne=119,720$ elements, and a third stiffly stable time advancement scheme is used with $\Delta t= 0.0001s$. For more details on the spectral/hp element method, see ~\cite{sherwin1999spectral}.

To generate the kernel, we solve the 1D model for $s$ simulation samples. To this end, we extract the vessels' centerlines and use them to create the 1D map of the abdominal aorta (shown in Figure \ref{Fig:AAmap}). Here, the inlet velocity is adapted from  \cite{maier1989human} for 3D simulation. To define this velocity profile in the form of Eq. \ref{inlet} we set $T=1s$ and $n_i=4$ and randomized parameter are listed in Table \ref{table:rand2}. The specifications of the problem are detailed in Table \ref{table:Geo2} in Appendix \ref{sec:appendix}, where the numbers of the vessels correspond with those in Figure \ref{Fig:AAmap}. In this problem outflow and cross-section areas are not randomized. Since we use the incompressible form of Navier-stocks equations, the cross-section area does not change in time; therefore, in 1D simulation, $\beta$ is set to a large value to maintain a $dA/dt \approx 0$. The base cross-sectional area, $A_0(x)$, is derived by fitting polynomials to MRI cross-sectional data and is the same for all 1D samples. Figure \ref{Fig:inletex2} shows the velocity inlet and its randomized profiles. We simulate $s=150$ samples and create the kernel with $r=41$ for which $\sum_{i=1}^r\sigma^2_i/\sum_{i=1}^s\sigma^2_i > 0.99$.

We demonstrate the accuracy of the GP model predictions across two scenarios: Case 1, where the measurements originate from 3D simulation, and Case 2, where the measurements are derived from 1D simulation. This specific setup aims to highlight the GP model's flexibility in adapting to various sources of input measurements. The locations of the measurements and predictions are depicted in Figure \ref{Fig:Geometryex2}. In this figure, the measurements represent the cross-sectional-averaged velocity at the inlets of the vessels, and the predictions are made at the midpoint of selected vessels. The chosen model of the abdominal vasculature includes 17 vessels, yet only time-series measurements at two spatial locations are utilized for all GP predictions in this scenario. This setup serves as an example of blood flow reconstruction in a data-poor regime.

In Case 1, we utilize average velocity data from 3D simulation as measurements $(\Delta t=0.001s)$, whereas in Case 2, the measurements are derived from 1D simulation $(\Delta t=0.005s)$. In Figure \ref{Fig:Abdresult_3D}, the predictions of the GP model for Case 1 and the ground truth, i.e., the cross-section-averaged velocity obtained from the 3D simulation, are shown. The GP model accurately predicts velocity in different vessels across various velocity ranges, despite no measurements being provided near point 3 and 4. This demonstrates that although the kernel is constructed with 1D simulations, it can effectively predict the results of 3D simulations, utilizing 3D time series data provided at only two locations.

In Case 2, the errors are significantly smaller compared to Case 1. This outcome is not surprising because, in Case 2, the ground truth is based on the 1D simulation results, and the kernel is constructed using 1D results as well. A comparison of the GP predictions between Cases 1 and 2, for example, Figure \ref{Fig:Abd5U} versus Figure \ref{Fig:Abd5U1d}, reveals differences in maximum velocities. Nonetheless, the GP model, based on the same physics-informed kernel, can accurately predict the velocity for both cases.


\subsection{Circle of Willis Vasculature}
\begin{figure}[htbp!]
	\centering
	\subfigure[3D-ToF]{
\raisebox{0.5cm} {\includegraphics[trim=0cm 0cm 0cm 0cm, clip=true, width=0.27\textwidth]{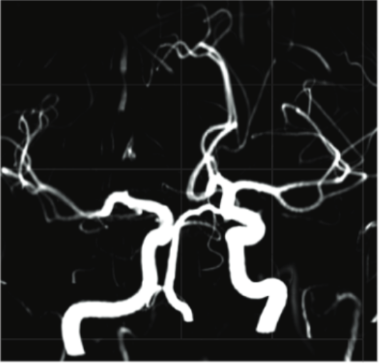}}
\label{Fig:3D-ToF}
}
	\subfigure[4D Flow MRI]{
\raisebox{0.5cm} {\includegraphics[trim=0cm 0cm 0cm 0cm, clip=true, width=0.27\textwidth]{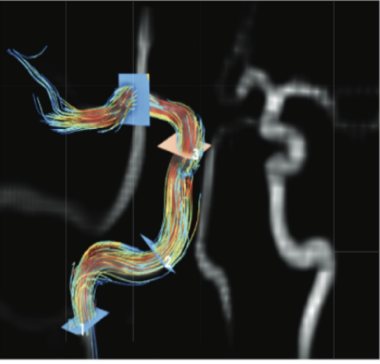}}
\label{Fig:4Df}
}
\subfigure[Schematic and point locations]{
\includegraphics[trim=0cm 0cm 0cm 0cm, clip=true, width=0.33\textwidth]{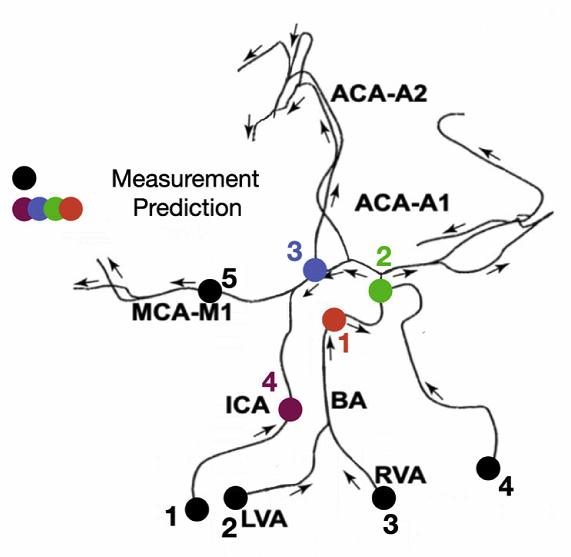}
\label{Fig:BrainMap}
}

	\caption{MRI data: a) 3D ToF MRI scan, representing the CoW architecture. b) 3D blood velocity streamlines at peak systolic phase through the left internal carotid artery acquired from processing the 4D Flow MRI images. c) The location of utilized measurements and predictions on extracted centerlines of the CoW arteries. Left and right internal carotid arteries are marked.}
	\label{fig::fig2}
\end{figure}

We predict blood flow velocity in the Circle of Willis (CoW) vasculature based on MRI measurements provided in \cite{sarabian2022physics}. In this reference, MRI data includes \textit{in vivo} blood flow velocity data across the CoW of a healthy male volunteer (age = 30 years, weight = $94 kg$, height = $185 cm$), which are collected as follows. First, we performed a ToF angiography scan, which is routinely performed in the clinic, is used to build the geometrical features of the CoW. In addition, we obtained a 4D Flow MRI scan, which provides 3D velocity time series maps in the entire CoW vasculature. The spatial resolution of 4D Flow MRI is $1.5\times 1.26\times 2$ $mm^3$, and the temporal resolution is 42 msec over 14 cardiac phases. All the MRI scans were performed on a 3T Siemens Skyra. The 3D-ToF scan, representing  the  CoW architecture, is depicted in Figure \ref{Fig:3D-ToF}. The vessel centerlines and branching pattern are shown in Figure \ref{Fig:BrainMap}.

In the processing of 4D Flow MRI images, we average the velocity data over a cross-sectional area to obtain 1D equivalent values. This is achieved by integrating the velocity components ($x$, $y$, and $z$) across each voxel in the imaging plane, and dividing by the total number of voxels in that plane. This spatial averaging procedure yields a representative velocity profile for the entire cross-section, effectively reducing the dimensionality of the data and enabling straightforward analysis and visualization of blood flow patterns. This analysis effectively produces similar values as TCD measurements at each specific cross section.

According to  \cite{sarabian2022physics}, the arterial geometry of the CoW is acquired by measuring 200 slices of $0.25mm$ thickness with a 3D-TOF MRI sequence. The in-plane resolution was $0.4 mm  \times 0.4mm$, the repetition time (TR) and the echo time (TE) were $18 ms$ and $3.57 ms$, respectively, and the flip angle was 15\textdegree. In the same scanning session, a prospectively ECG-gated 4D Flow MRI scan was performed using a 3.0 Tesla MRI scanner (Skyra, Siemens Healthcare, Erlangen, Germany) and a 32-channel head coil. Scanning parameters were as follows: flip angle 7\textdegree, repetition time/echo time $44.56/2.78 ms$ , bandwidth $445$ Hz/pixel, velocity encoding (VENC) $90 cm/s$, voxel size $1.5\times 1.26 \times 2 mm^3$, temporal resolution  $41.36 ms$ over $14$ cardiac phases.

Before extracting velocity from 4D Flow MRI phase difference images, we perform image pre-processing including noise-masking, eddy-currents correction as well as phase unwrapping following the method explained in \cite{bock2007optimized}. Noise masking was performed by thresholding of the signal intensity in the magnitude data to exclude regions with low signal intensity. Eddy current correction consisted of three steps: i) separation of static regions from blood flow, ii) fit a plane with least-squares method to the static regions from the last time frame (late diastole), iii) Subtraction of the fitted plane to the MRI data in every time frame. Finally, we correct the velocity aliasing effect, \textit{i.e,} if adjacent pixels velocities in temporal or slice direction differ by more than VENC. 


As described in \cite{sarabian2022physics}, the 3D and 1D geometry of the vasculature is extracted from the 3D-TOF slices using SimVascular \cite{updegrove2017simvascular}. Where, the threshold for the basic segmentation was set to one-fourth of the maximum signal intensity value obtained in the measurements, which showed the best results for segmenting arterial structures. Everything above this value was considered to be an artery. Additional segmentation was performed manually by adding or removing voxels from the geometry based on anatomical knowledge.

Here, we use 1D geometry and properties used in \cite{sarabian2022physics} to perform 1D simulations. This problem has four inlets where the direction of the blood flow is shown in Figure \ref{Fig:BrainMap}. The cardiac cycle is $T=0.579s$ and each inlet is defined using Eq. \ref{inlet} and randomized as listed in Table \ref{table:randBrain}. The inlets are located at the same positions as measurement points 1, 2, 3, and 4.

{\small
\begin{table}[H]
\caption{Randomized parameters for CoW.}
\centering
\resizebox{\textwidth}{!}{%
\begin{tabular}{| c | c | c | c | c | c | c | c | c |}
\hline
Inlet & j & k & $a_0$ & $a_i$ & $b_i$ & $c_i$\\ [0.5ex] 
\hline
\multirow{2}{*}{1} & \multirow{2}{*}{6} & $\overline{\xi}_k$ & 0.06 &  [0.08,3.6e-2,0.23,0.01,0.02,-8e-3]& [0.1,0.24,0.4,0.05,0.5,0.02] & [4e-3,1.5e-2,9e-3,1e-3,3e-3,1.4e-3]  \\
\cline{3-7}
 & & $\sigma_k$& 0.08 & 0.08 & 0.05 & 1e-4 \\
\hline
\multirow{2}{*}{2,3} & \multirow{2}{*}{5} & $\overline{\xi}_k$ & 0.08 &  [0.09,6.5e-2,0.04,1.2e-2,-0.03]& [0.1,0.2,0.36,0.5,0.03] &[1e-3,0.02,9e-4,2e-3,8e-4]  \\
\cline{3-7}
 & & $\sigma_k$& 0.08 & 0.08 & 0.05 & 1e-4 \\
\hline
\multirow{2}{*}{4} & \multirow{2}{*}{5} & $\overline{\xi}_k$ & 0.08 &  [0.06,0.03,1.4e-2,-0.02,-1.5e-2]& [0.08,0.2,0.4,0.6] & [2e-3,5e-3,1.5e-2,0.01]  \\
\cline{3-7}
 & & $\sigma_k$& 0.08 & 0.08 & 0.05 & 1e-4 \\
\hline
\end{tabular}
}
\label{table:randBrain}
\end{table}
 }
We also randomize $R_t$ and $C$ using Eq. \ref{area_outlet} where $\sigma_{R_t} = \sigma_C = 0.05$. $A_0(x)$ is determined by fitting polynomials to cross-sectional data obtained from MRI scans and we randomize the area using $\sigma_A=0.5$. Using these parameters, we generate $s=150$, 1D samples to create the data matrix (Eq. \ref{DataMatrix}). We use $r=36$ which captures more than 95\% variance of the data.

\begin{figure}[t!]
\centering
\subfigure[Point 1, MRI]{
\includegraphics[trim=2cm 0cm 1.5cm 0cm, clip=true, width=0.22\textwidth]{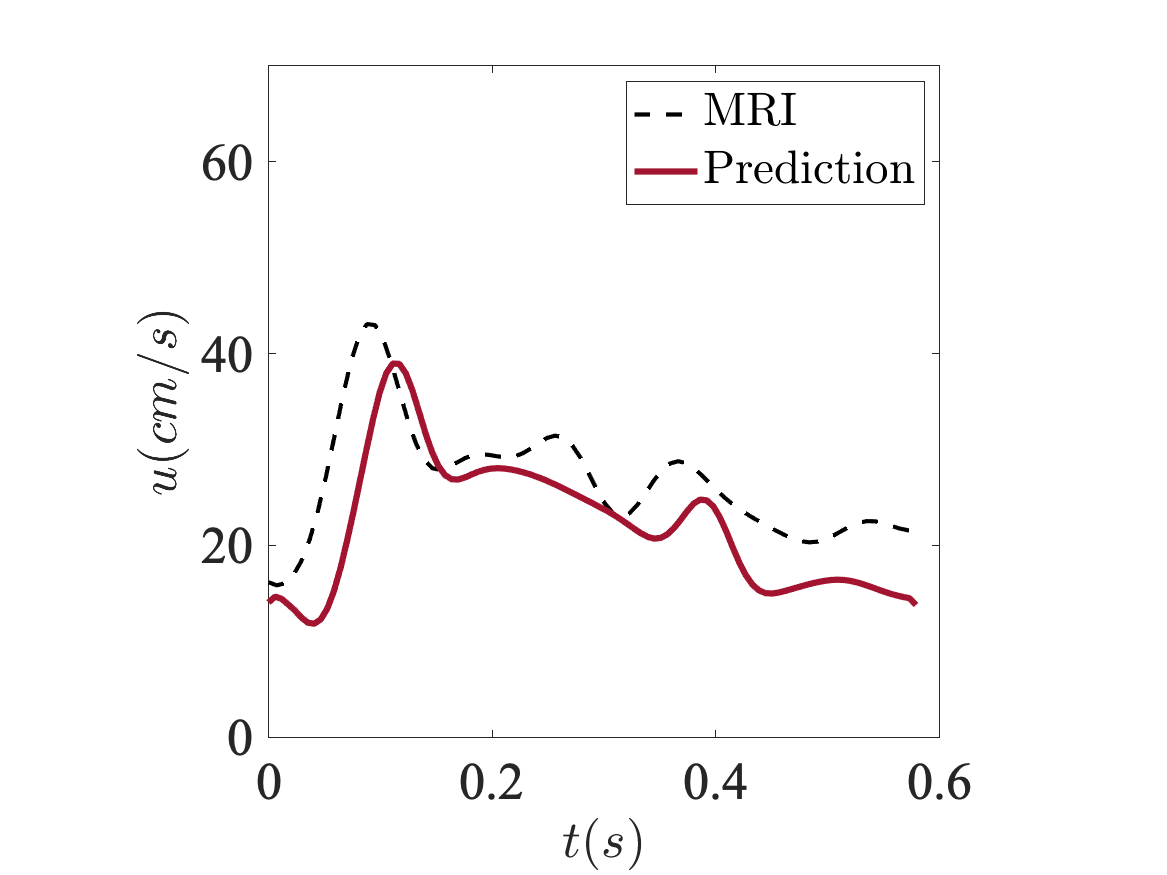}
\label{Fig:MRIp1}
}
\subfigure[Point 2, MRI]{
\includegraphics[trim=2cm 0cm 1.5cm 0cm, clip=true, width=0.22\textwidth]{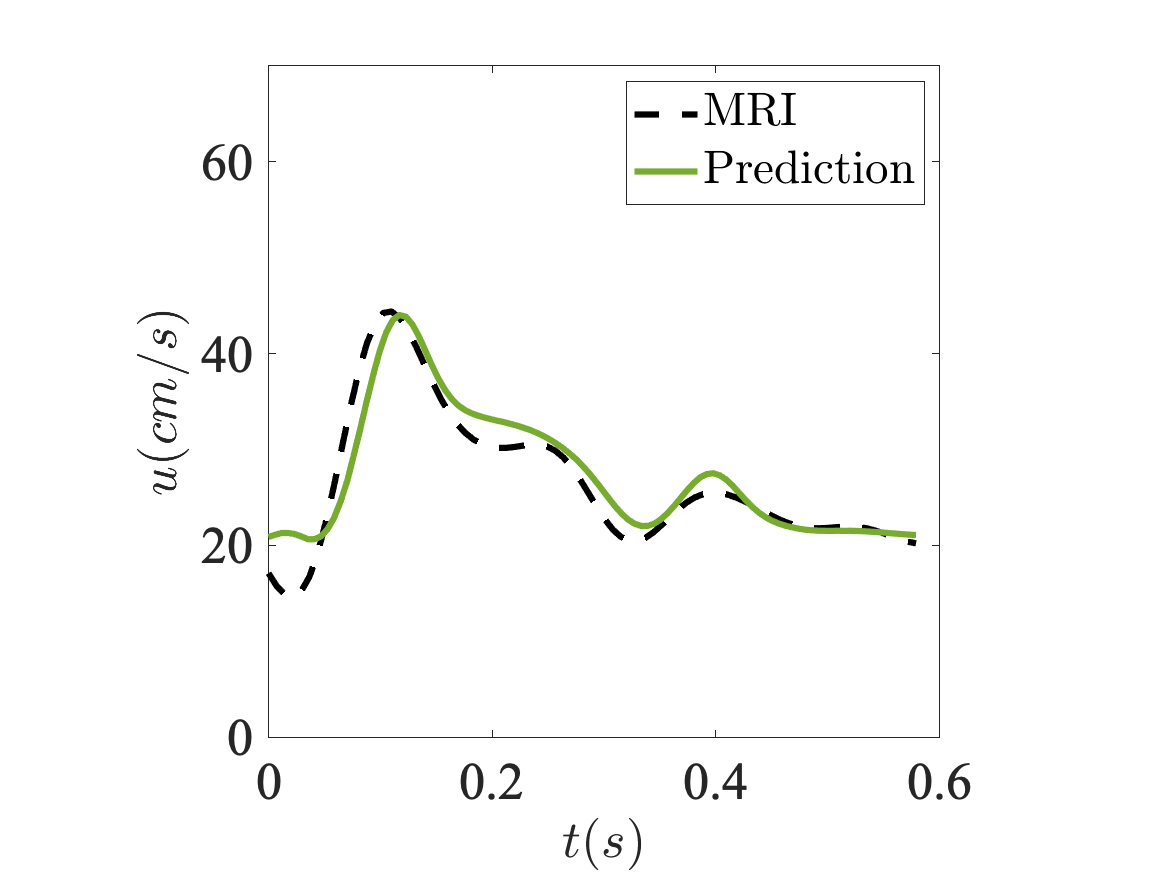}
\label{Fig:MRIp2}
}
\subfigure[Point 3, MRI]{
\includegraphics[trim=2cm 0cm 1.5cm 0cm, clip=true, width=0.22\textwidth]{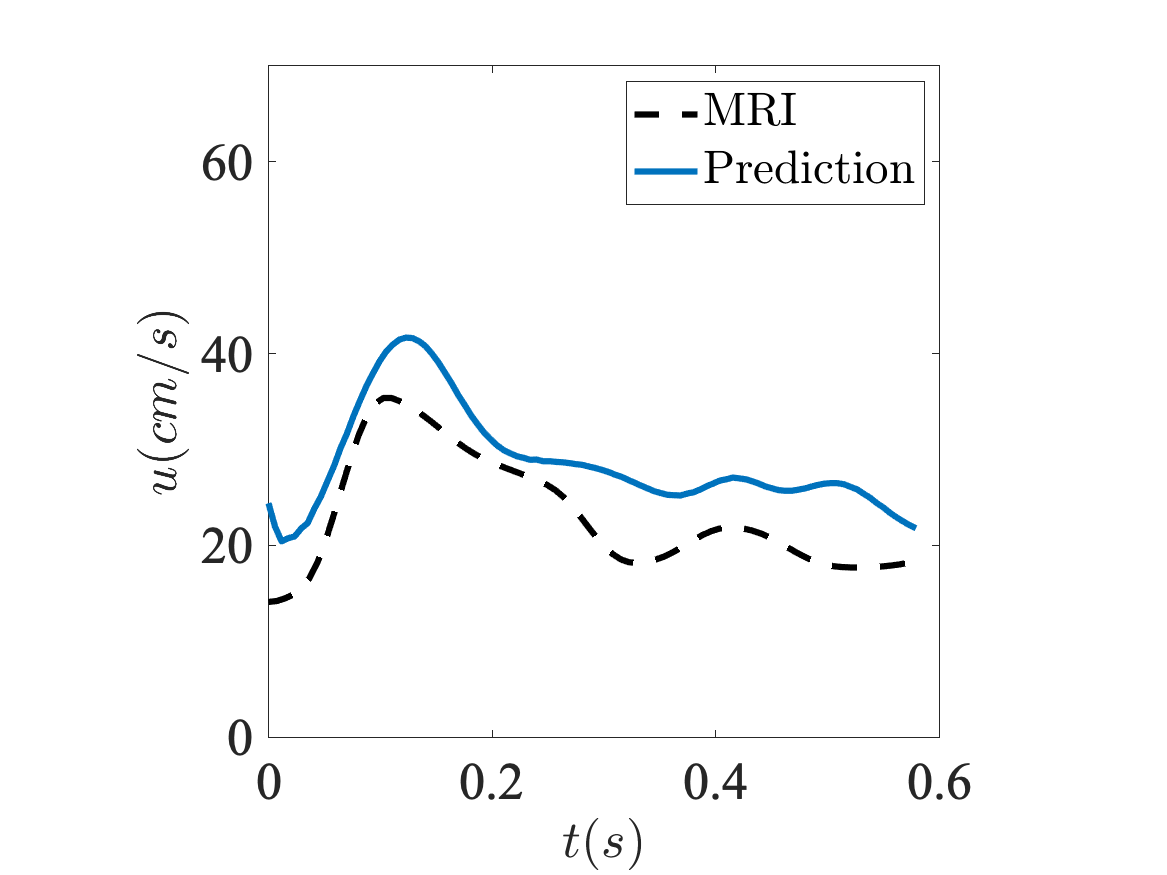}
\label{Fig:MRIp3}
}
\subfigure[Point 4, MRI]{
\includegraphics[trim=2cm 0cm 1.5cm 0cm, clip=true, width=0.22\textwidth]{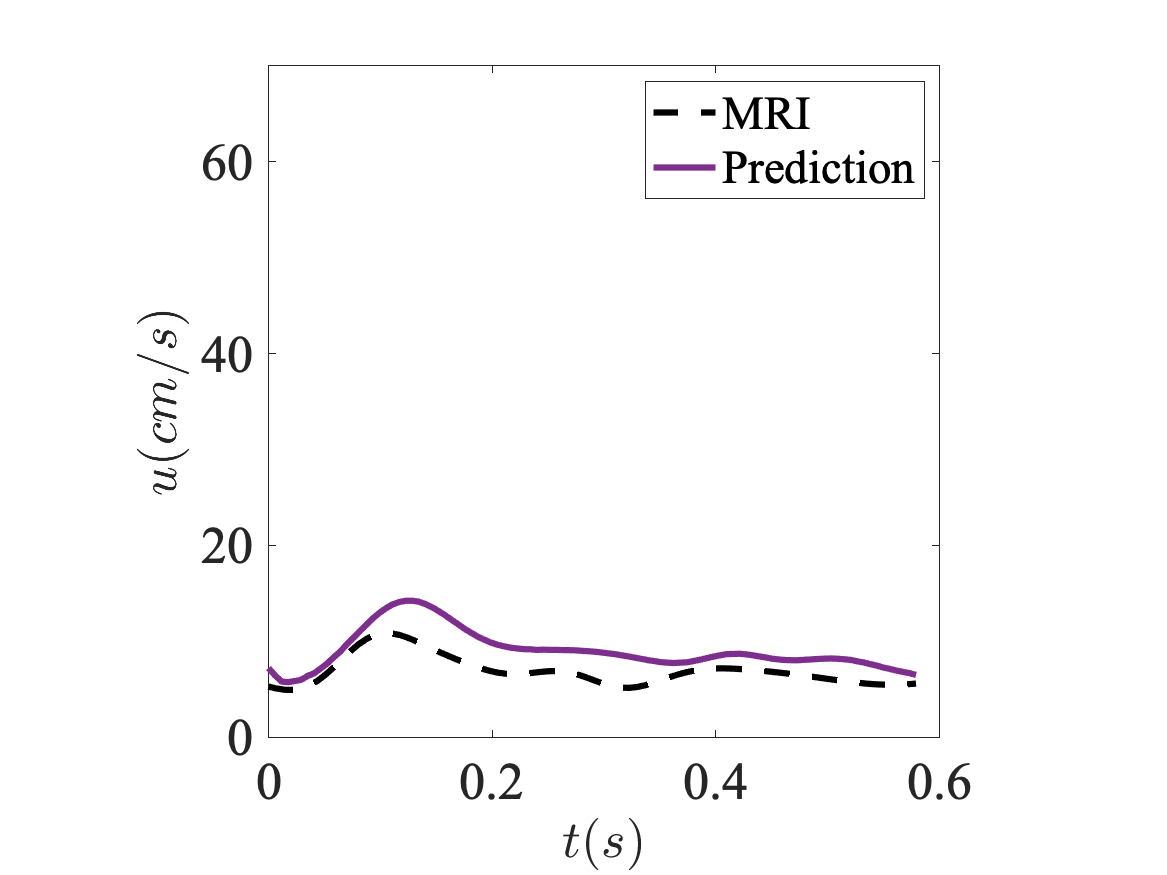}
\label{Fig:MRIp4}
}
\caption{Case 1: Prediction of the brain vasculature using MRI measurements. The location of these points are provided in Figure \ref{Fig:BrainMap}. Predictions are compared with 4D Flow MRI data at a) point 1, b) point 2, c) point 3, and d) point 4. The observed discrepancy is due to measurement errors related to mass conservation, whereas the GP model inherently enforces mass conservation in its predictions.}
\label{MRIResult}
\end{figure}
\begin{figure}[t!]
\centering
\subfigure[Point 1, 1D]{
\includegraphics[trim=2cm 0cm 1.5cm 0cm, clip=true, width=0.22\textwidth]{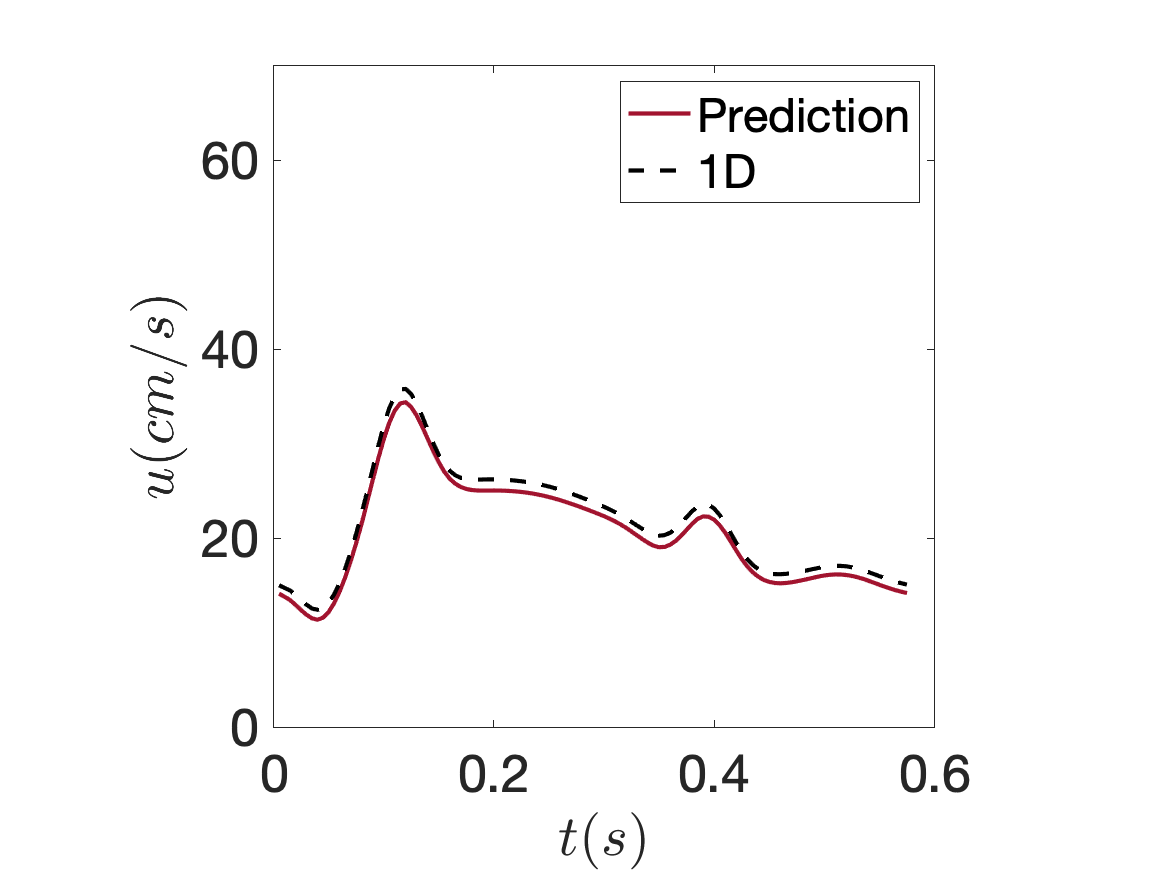}
\label{Fig:MRI1dp1}
}
\subfigure[Point 2, 1D]{
\includegraphics[trim=2cm 0cm 1.5cm 0cm, clip=true, width=0.22\textwidth]{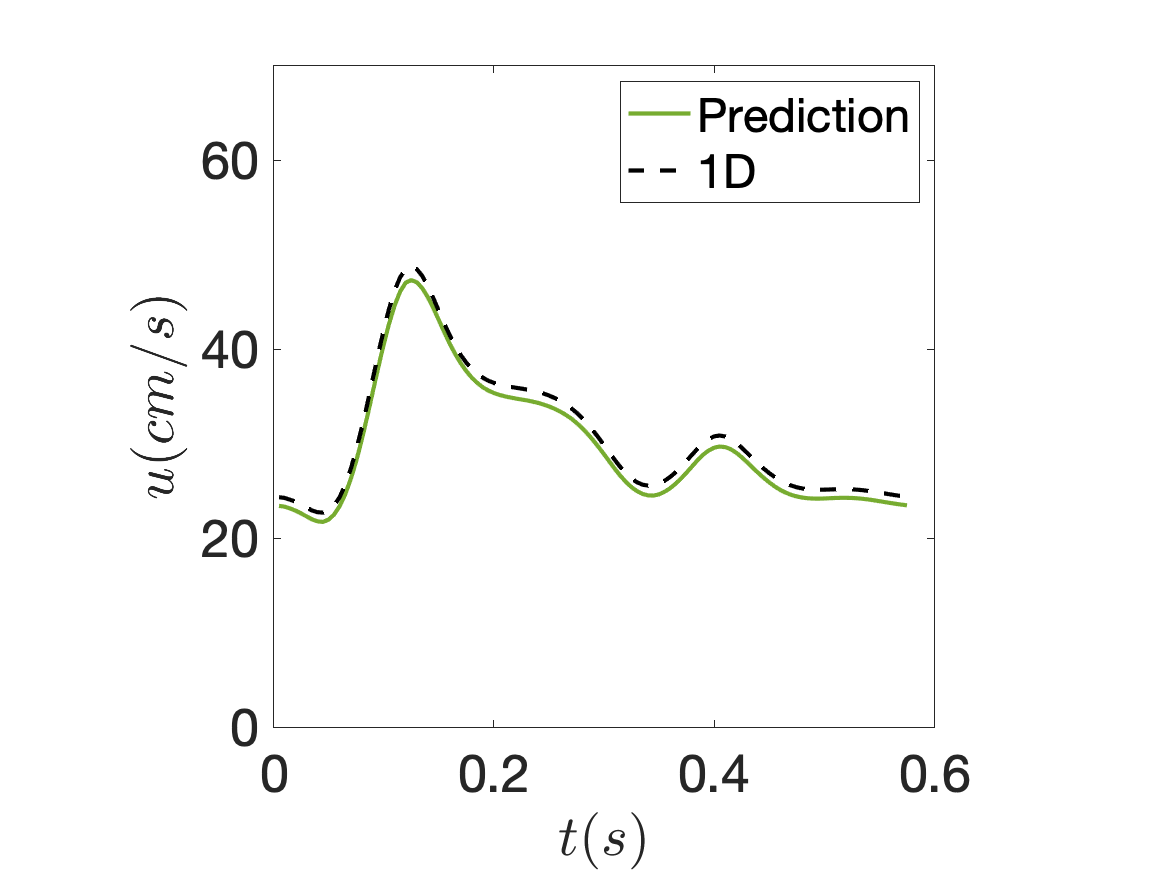}
\label{Fig:MRI1dp2}
}
\subfigure[Point 3, 1D]{
\includegraphics[trim=2cm 0cm 1.5cm 0cm, clip=true, width=0.22\textwidth]{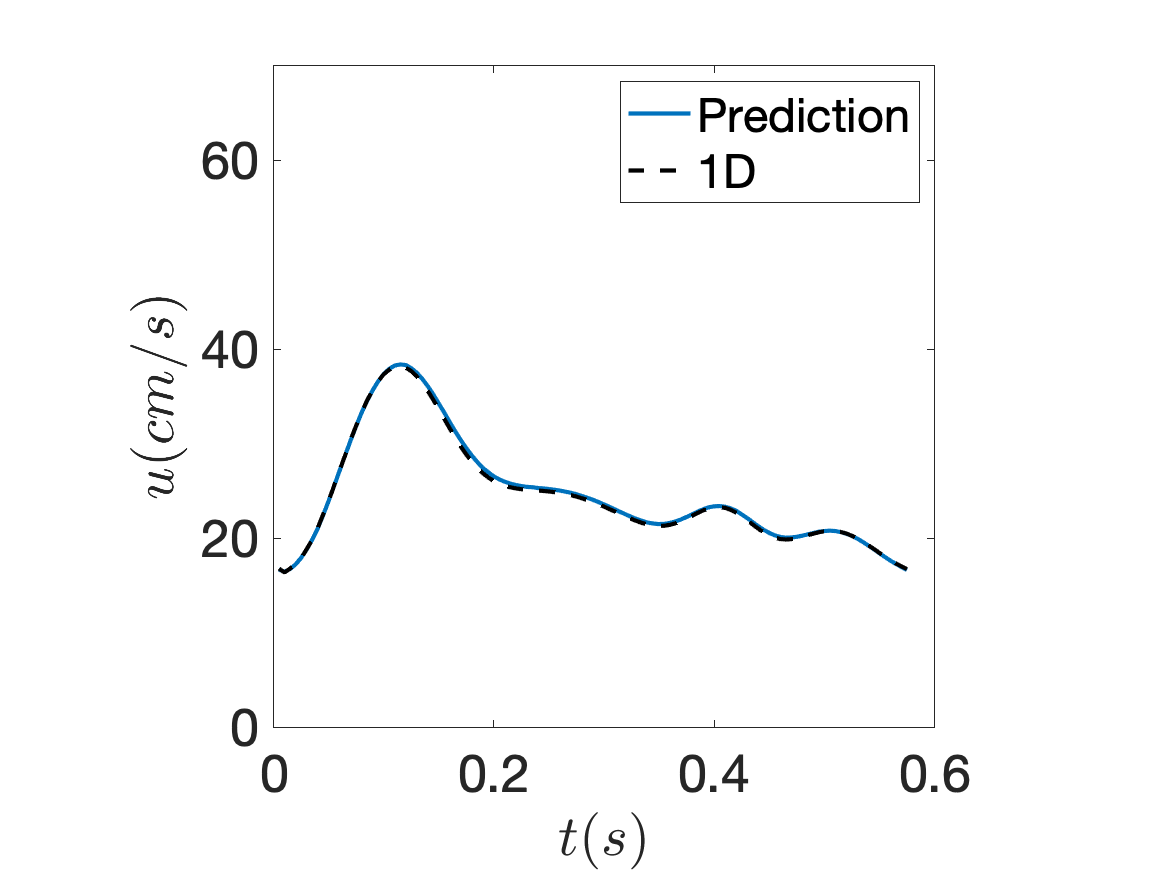}
\label{Fig:MRI1dp3}
}
\subfigure[Point 4, 1D]{
\includegraphics[trim=2cm 0cm 1.5cm 0cm, clip=true, width=0.22\textwidth]{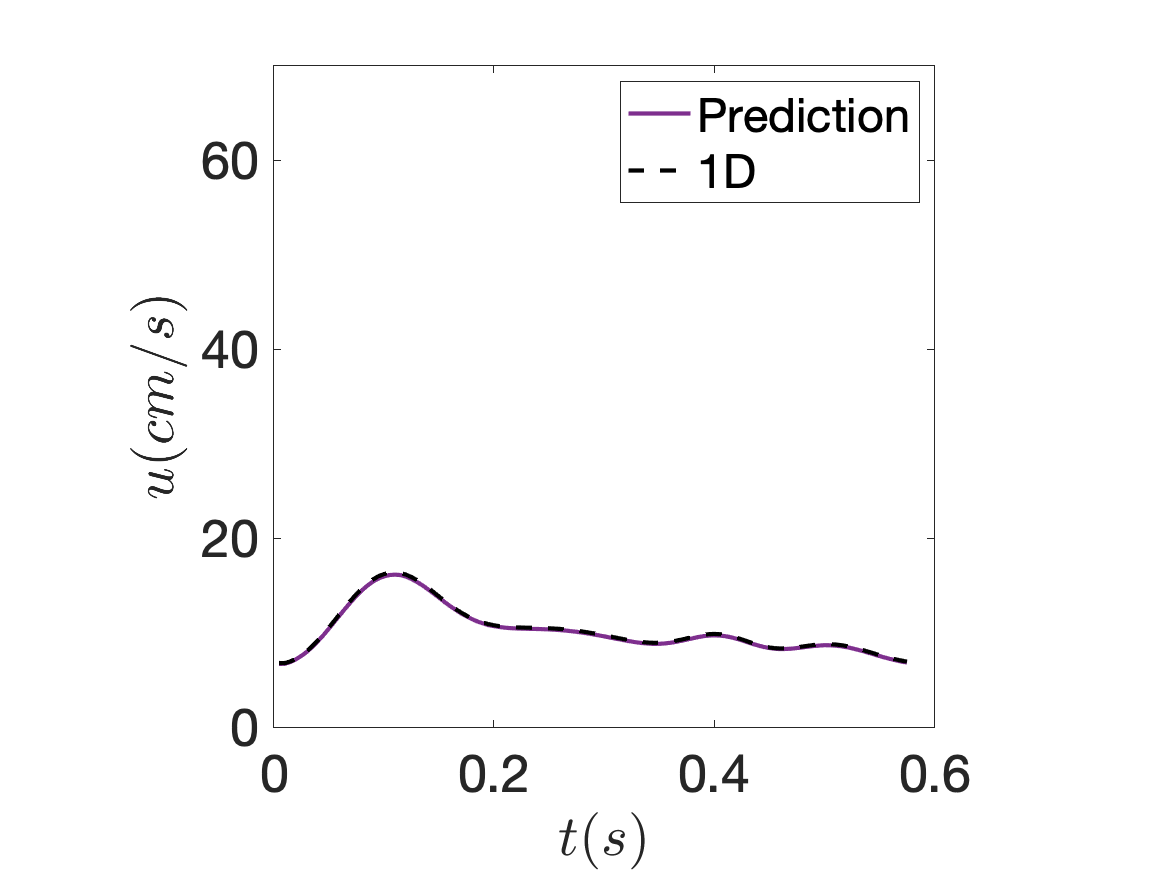}
\label{Fig:MRI1dp4}
}
\caption{Case 2: Prediction of the brain vasculature using 1D simulation data as measurements. Predictions are compared with 1D simulations at a) point 1, b) point 2, c) point 3, and d) point 4 which their locations are indicated in Figure \ref{Fig:BrainMap}. Based on these comparison, GP predictions are in good agreement with 1D simulations across different vessels.}
\label{MRIResult_1d}
\end{figure}

We examine predictions for two scenarios: In Case 1, we utilize the collected 4D Flow MRI data (average velocity) as measurements; while in Case 2, we use 1D simulation data taken at the same locations  as measurements. Measurements locations are shown in Figure \ref{Fig:BrainMap} where measurements points 1-4 are located at the inlets and measurement point 5 is at the middle of the vessel. Figure \ref{MRIResult} compares the prediction with average velocity from 4D Flow MRI (labeled MRI for brevity) in Case 1. We also compare the GP predictions with 1D simulations for Case 2 as shown in Figure \ref{MRIResult_1d}. In Case 2, where the measurement points are taken from 1D simulation, the accuracy is very high. However, in Case 1, where the ground truth is 4D Flow MRI, we observe around 20\% inaccuracy. After closely inspecting the 4D Flow MRI data, it was realized that the 4D Flow MRI measurements violate the conservation of mass up to 27\%. On the other hand, the GP predictions by construction satisfy the conservation of mass. 

\begin{figure}[t!]
\centering
\subfigure[Mean convergence]{
\includegraphics[clip=true, width=0.4\textwidth]{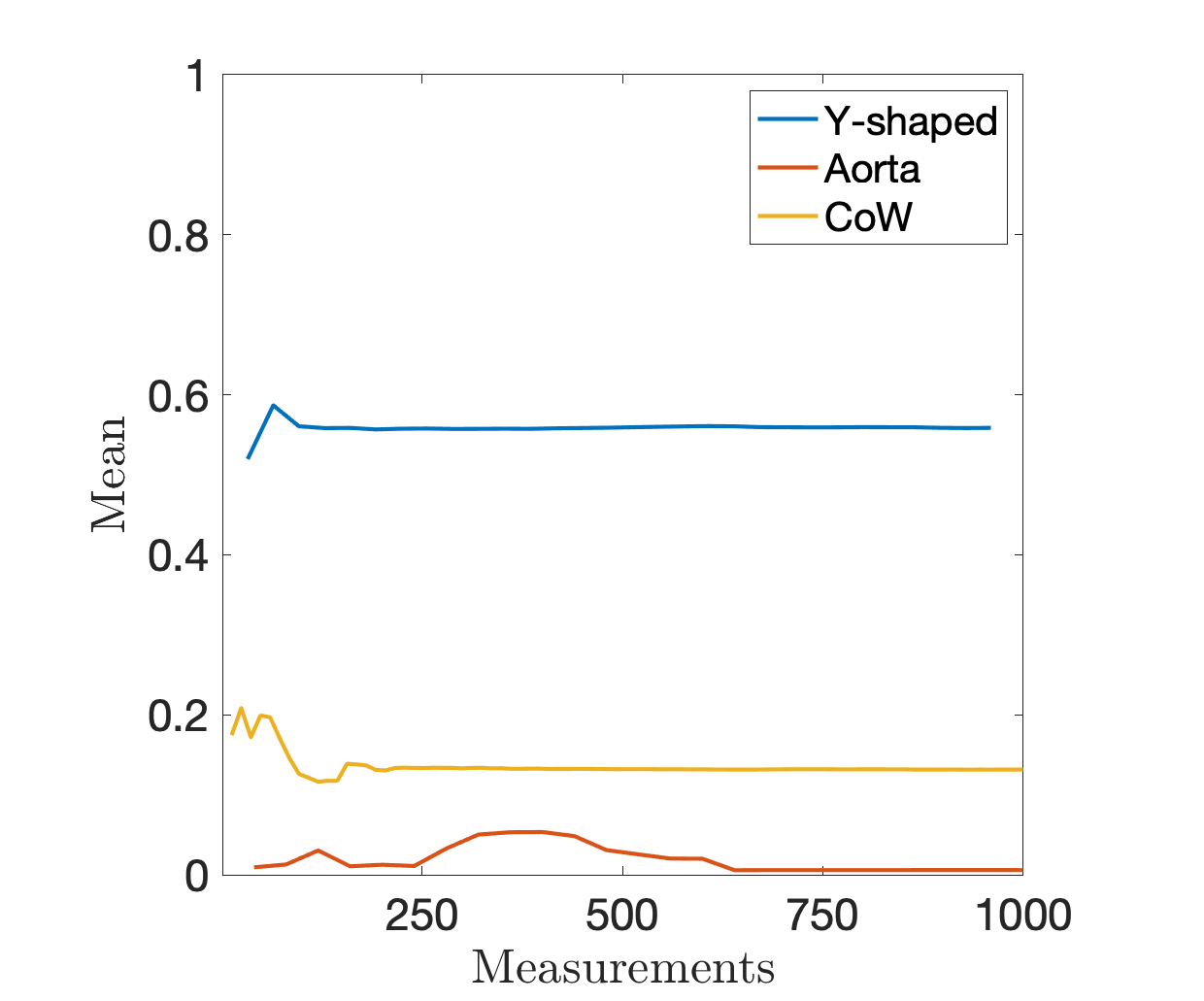}
\label{Fig:Brain_mean}
}
\subfigure[Uncertainty convergence]{
\includegraphics[clip=true, width=0.4\textwidth]{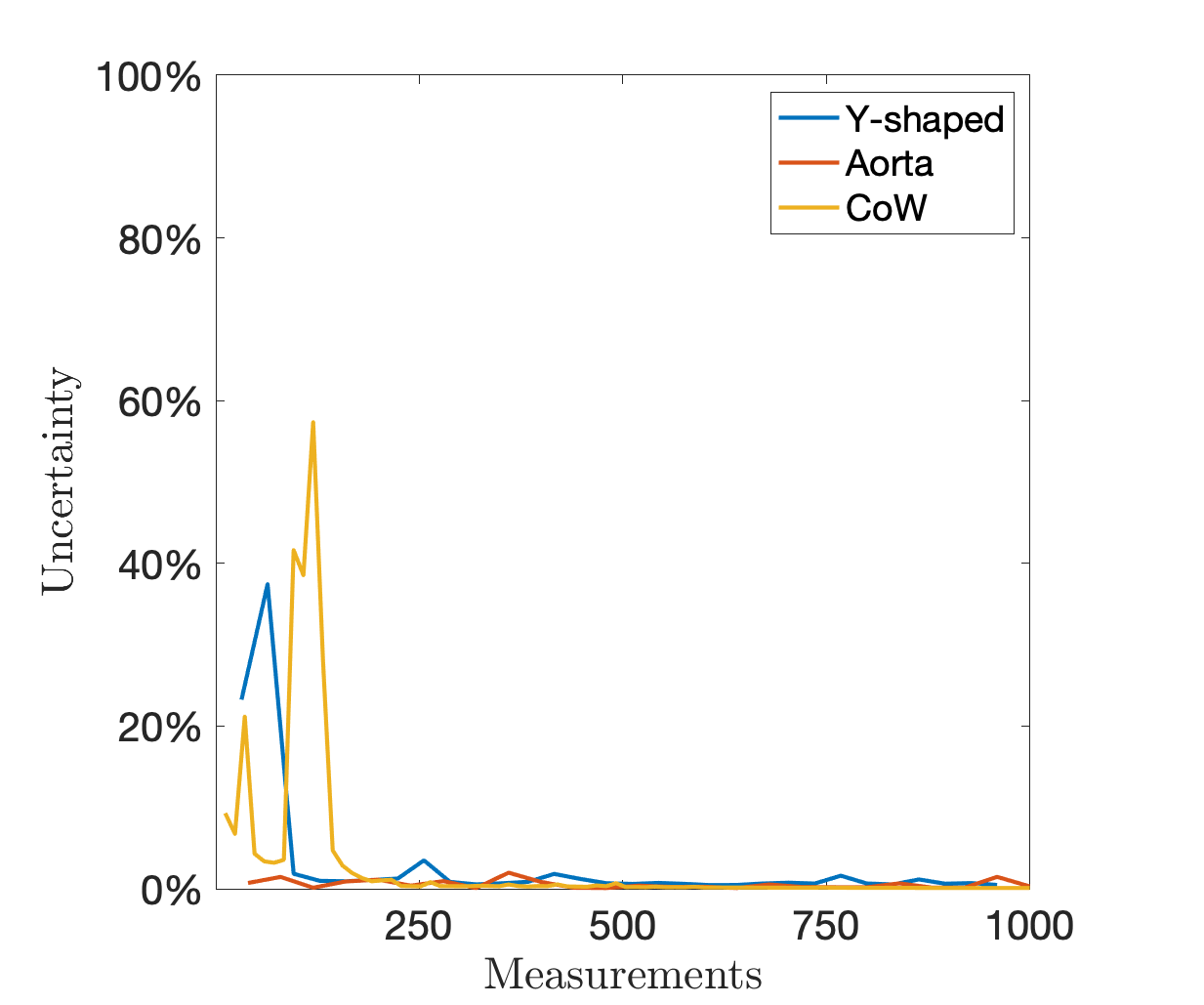}
\label{Fig:Brain_sigma}
}

\caption{a) Average mean convergence vs number of utilized measurements. b) Average uncertainty percentage convergence vs number of utilized measurement.}
\label{Fig:sigma_mean_conv}
\end{figure}

The posterior uncertainty in both Cases 1 and 2 is large (not shown in the figures) due to the lack of having very small amount of data compared to the size of the problem.  In Figure \ref{Fig:sigma_mean_conv}, we show the convergence of the spatiotemporal averaged mean and standard deviation of the GP mean versus the number of measurement points in time and space. In the figures, the mean axis indicates the average mean across all prediction points, while the uncertainty percentage axis shows the average standard deviation to mean ratio for all prediction points. These results show that as more data become available the GP mean converges and the uncertainty reduces.



\section{Conclusions}\label{sec:Disc}
We introduce a novel methodology that enables the spatiotemporal reconstruction of blood flow velocity within a vasculature network from a relatively small number of observations. This approach is particularly relevant for clinical scenarios such as blood flow reconstruction using TCD ultrasound measurements, which can only capture the blood flow velocity time series at a few points. Our methodology leverages GP regression and introduces a novel approach to construct a spatiotemporal kernel. This kernel is formulated using data generated from 1D blood flow simulations with a set of random parameters. These parameters encapsulate the epistemic uncertainties, such as the lack of precise knowledge about boundary conditions and the geometry of the vessels, for instance, the cross-sectional areas of the blood vessels. One significant feature of the constructed kernel is its encoding of vessel-to-vessel correlations, allowing measurements from one vessel to approximate velocities in others. 
Also, any prediction using the constructed kernel satisfies the conservation of mass. 

We demonstrate the performance of the presented methodology through three case studies: (i) a Y-shaped vessel, investigating the impact of measurement placement and their spatial and temporal resolution on accuracy and uncertainty bounds; (ii) abdominal aorta, utilizing a 3D model extracted from MRI scans with its simulation data as measurements. For comparison, we also examine a scenario where the 3D simulation data is replaced by 1D simulation data. In both scenarios, data from only two locations are used to accurately predict blood flow velocity in all 17 vessels; (iii) CoW, employing 4D Flow MRI as the measurement data and exploring prediction results when 1D simulation data is used instead of MRI measurements.

To put things in perspective, the presented algorithm seeks to find a low-dimensional spatiotemporal function space to approximate velocity, i.e., \(u(x,t) = \sum_{i=1}^r w_i \phi_i(x,t)\), as opposed to using fixed basis functions, i.e., \(u(x,t) = \sum_{i=1}^{n_x}\sum_{j=1}^{n_t} w_{i,j} \psi_i(x) \chi_j(t)\), which requires inferring \(\mathcal{O}(n_v n_x n_t)\) parameters, where \(n_v\) is the number of vessels. However, \(r \ll n_v n_x n_t\), meaning that the presented methodology requires the inference of \(r\) parameters, and therefore, it requires a significantly smaller amount of data. PINN also seeks to find \(r\) adaptive spatiotemporal basis functions and approximate velocity with \(u(x,t) = \sum_{i=1}^r w_i \phi_i(x,t;\theta)\). However, a PINN requires solving a nonconvex optimization problem to find \(\theta\) and the \(w_i\)'s, and it requires data on pressure, which is not clinically available. As a result, the computational cost of constructing the kernel is significantly lower than that of PINNs.
Specifically, the offline (training) cost involves running $\mathcal{O}(100)$ 1D blood flow simulations, each taking only minutes for vasculature with $\mathcal{O}(100)$ vessels. Moreover, these simulations can run concurrently since they are independent of each other. The computation of the kernel, involving an SVD on the data generated by these simulations, costs about $\mathcal{O}(10)$ seconds and is negligible compared to the total cost. The online (inference) cost is minimal because the number of observations is typically very small, allowing for the methodology to be used for real-time flow reconstruction.

Both offline and online computational costs of the presented methodology scale linearly with the size of the problem, i.e., the number of vessels. More broadly, the presented methodology can be similarly developed and utilized for other network-type problems, such as fluid flows in pipe networks, traffic flow, or power grids.

\appendix
\section{Appendix}\label{sec:appendix}
\subsection{One-Dimensional Simulation}\label{sec:physics}
\hspace{0.1in}The governing equations for 1D simulations are the simplified version of the Navier-Stokes equations based on the following assumptions: i) The vessel's curvature is small and negligible; hence the given equations are written based on $x$ coordinates on the centerline. ii) Vessels are axisymmetric. iii) The properties of each vessel are constant. These assumptions simplify the 3D incompressible Navier-Stokes equations which are written as:
     
   \begin{equation}
    \frac{\partial A}{\partial t} + \frac{\partial (Au)}{\partial x} = 0
    \label{conMass1d},
   \end{equation}
    \begin{equation}
    \frac{\partial u}{\partial t} + u \frac{\partial u}{\partial x} = - \frac{1}{\rho} \frac{\partial p}{ \partial x}  + \frac{f u}{ A}, \mbox{where}  \quad 
 f=-22 \mu \pi
    \label{Consmoment1d}.
   \end{equation}
$\rho=1050 kg/m^3$, $\mu$, $A(x,t)$, $p(x,t)$, $f(x,t)$ are the blood density, blood dynamic viscosity, vessel cross-section area, pressure, and friction force per unit length respectively. Here, $A$, $u$, and $p$ are unknown. In order to have a closed system of equation, we use pressure area relation.
   \begin{equation}
    p=p_{ext}+ \beta (\sqrt{A}-\sqrt{A_0}))
    \label{press1d}
   \end{equation}
   with
   \begin{equation}
   \beta = \frac{\sqrt{\pi}hE}{(1-\nu^2)A_0}
    \label{beta}
   \end{equation}
where $p_{ext}$, $h$, $E$, $A_0$, and $\nu$ are external pressure, wall thickness, Young modulus, equilibrium cross-section area, and Poisson ratio $\nu = 0.5$ respectively. To compute 1D blood flow properties, characteristic decomposition is used for the given equations. This derivation is described in detail in Ref. \cite{sherwin2003one} and solved by spectral/hp element spatial discretization and a second-order Lax-Wendroff time-integration scheme. This solver is called Nektar and we use it to solve $s$ different simulations/samples with randomized inlet velocity, cross-section areas, and outflow boundary conditions. 

\subsection{Abdominal Aorta}
The following table provides abdominal aorta specifications for 1D simulations.
\begin{table}[H]
\caption{Abdominal aorta vasculature 1D simulation specification}
\centering
\begin{tabular}{| c | c | c | c | c | c |}
\hline
Vessel & Length$(m)$  & Compliance$(10^{-10}\frac{m^3}{Pa})$ & Total resistance$(10^{10}\frac{Pa.s}{m^3})$ & $\beta(\frac{Pa}{m})$\\ [0.5ex] 
\hline
\rule{0pt}{4ex} 
1&  0.0346 & - & - & 1.083e9\\
2&  0.0546 & $C_1=$0.0102 & $R_{t_1}=$4.789 & 1.0522e10\\
3&  0.0578  & - & - & 1.0396e10\\
4&  0.077&  - & - & 3.2512e9\\
5&  0.1701&  $C_2=$0.1931 & $R_{t_2}=$0.6989 & 1.0396e10\\
6&  0.0285&  - & - & 3.1355e9\\
7&  0.0489&  $C_3=$0.0936 & $R_{t_3}=$5.35 & 5.2091e10\\
8&  0.0152& - & - & 5.2163e10\\
9&  0.0101& $C_4=$0.0194 & $R_{t_4}=$4.1823 & 1.0332e11\\
10& 0.0217& $C_5=$0.0147 & $R_{t_5}=$3.2823 & 1.0274e10\\
11& 0.0414&  - & - & 5.1574e9\\
12& 0.1994& $C_6=$0.0629 & $R_{t_6}=$0.5974 & 6.277e9\\
13& 0.0212& - &  - & 5.4503e9\\
14& 0.0431&  $C_7=$0.011 & $R_{t_7}=$6.1527 & 2.8957e10\\
15& 0.0138& - & - & 5.063e10\\
16& 0.029 & $C_8=$0.0439 &  $R_{t_8}=$4.5299 & 3.063e10 \\
17& 0.0118&  $C_9=$0.0128 & $R_{t_9}=$9.5299 & 5.063e10\\
\hline
\end{tabular}
\label{table:Geo2}
\end{table}

\section*{Acknowledgements}
This project was supported by the National Institutes of Health, Grant No. R21EB032187 and partially by the Air Force Office of Scientific Research award No. FA9550-21-1- 0247.
\enlargethispage{20pt}








\bibliographystyle{ieeetr}

\newpage

\end{document}